\documentclass{article}

\usepackage{arxiv}

\usepackage[utf8]{inputenc} 
\usepackage[T1]{fontenc}    
\usepackage{hyperref}       
\usepackage{url}            
\usepackage{booktabs}       
\usepackage{amsfonts}       
\usepackage{nicefrac}       
\usepackage{microtype}      
\usepackage{lipsum}
\usepackage{float}
\usepackage{graphicx}
\graphicspath{ {./images/} }
\usepackage{subcaption}  
\usepackage{float}        
\usepackage{amssymb}
\usepackage[export]{adjustbox}
\usepackage{xcolor}
\usepackage{tcolorbox}
\usepackage{appendix}           
\usepackage{amsmath}
\usepackage{enumitem}
\usepackage{framed}
\usepackage{booktabs}  
\usepackage{tocloft}
\usepackage{multirow}
\usepackage{amsthm}
\usepackage{tikz}
\usepackage{caption} 
\usepackage{array}
\usepackage{tabularx}
\usepackage{booktabs}   
\usepackage{makecell}   
\renewcommand{\arraystretch}{1.4}
\usepackage[table]{xcolor} 
\definecolor{highlight}{RGB}{255,220,220} 

\usetikzlibrary{shapes.geometric, arrows.meta, positioning} 

\newtheorem{hypothesis}{Hypothesis}

\title{When Your AI Agent Succumbs to Peer-Pressure: \\Studying Opinion-Change Dynamics of LLMs}

\author{
  Aliakbar Mehdizadeh \\
  Department of Communication \\
  University of California, Davis \\
  \texttt{amehdizadeh@ucdavis.edu} \\
   \And
  Martin Hilbert \\
  Department of Communication \\
  University of California, Davis\\
  \texttt{hilbert@ucdavis.edu}
}

\begin{document}
\maketitle

\begin{abstract}
We investigate how peer pressure influences the opinions of Large Language Model (LLM) agents across a spectrum of cognitive commitments by embedding them in social networks where they update opinions based on peer perspectives. Our findings reveal key departures from traditional conformity assumptions. First, agents follow a sigmoid curve: stable at low pressure, shifting sharply at threshold, and saturating at high. Second, conformity thresholds vary by model: Gemini 1.5 Flash requires over 70\% peer disagreement to flip, whereas ChatGPT-4o-mini shifts with a dissenting minority. Third, we uncover a fundamental "persuasion asymmetry," where shifting an opinion from affirmative-to-negative requires a different cognitive effort than the reverse. This asymmetry results in a "dual cognitive hierarchy": the stability of cognitive constructs inverts based on the direction of persuasion. For instance, affirmatively-held core values are robust against opposition but easily adopted from a negative stance, a pattern that inverts for other constructs like attitudes. These dynamics echoing complex human biases like negativity bias, prove robust across different topics and discursive frames (moral, economic, sociotropic). This research introduces a novel framework for auditing the emergent socio-cognitive behaviors of multi-agent AI systems, demonstrating their decision-making is governed by a fluid, context-dependent architecture, not a static logic.\end{abstract}


\section{Introduction}

Understanding how values, beliefs, attitudes, opinions, and intentions evolve through collective influence is a central concern in communication science and social dynamics. The understanding the mechanisms that govern the formation and collective evolution of values, beliefs, attitudes, opinions, and intentions informs much of general social dynamics. This interdisciplinary endeavor seeks to unravel how individual interactions aggregate to produce emergent, macro-level social phenomena such as widespread consensus, persistent polarization, or societal fragmentation, and micro-level drivers such as the fundamental principles behind the diffusion of ideas, the survival of minority viewpoints, and the crystallization of social norms. 

Utilizing mathematical frameworks and agent-based computational models, historically, the field of opinion dynamics has been dominated by two major classes of models offering distinct lenses on social influence. The first, discrete opinion models, represents opinions as binary states (e.g., support/oppose) and draws inspiration from statistical physics, with foundational examples like the Voter Model, which captures social imitation~\cite{clifford1973model,holley1975ergodic}, and the Sznajd Model, based on the principle of "united we stand, divided we fall"~\cite{sznajd2000opinion}. The second tradition involves continuous opinion models, which represent opinions on a real-number scale and include the classical DeGroot Model, where agents average their neighbors' opinions~\cite{degroot1974reaching}, and the highly influential Bounded Confidence Models (such as Deffuant-Weisbuch and Hegselmann-Krause), which introduce a "confidence radius" to explain the emergence of stable opinion clusters and polarization by limiting interactions to agents with sufficiently similar views~\cite{deffuant2000mixing,hegselmann2002opinion}. In all the classical models, from the Voter Model~\cite{clifford1973model,holley1975ergodic} to the HK model~\cite{hegselmann2002opinion}, agents share a fundamental characteristic: they are rule-based. An agent's behavior is determined by a simple, pre-programmed mathematical function, such as "adopt the majority opinion of your neighbors," which serves as the basis for the Majority Vote Model (MVM) used as a baseline in this study~\cite{oliveira1992nonequilibrium}. This approach has been remarkably successful in generating complex macro-behavior from simple micro-motives~\cite{castellano2009statistical}. 

However, this rule-based methodological paradigm can, and should, now be expanded by a new class of agents. The research presented here marks a departure from simulating social dynamics with abstract, rule-following entities, using, instead reasoning, language-processing agents powered by Large Language Models (LLMs). These agents are not given a fixed behavioral rule; rather, they are presented with a natural-language description of their social context and are prompted to reason about their cognitive state (such as opinion, attitude, belief), which has been trained on unprecedented amounts of human interaction data. This represents a significant methodological shift in computational communication science, enabling a more flexible, more nuanced, but also less controllable exploration of social theories that have historically been difficult to reduce to simple mathematical formalisms~\cite{park2023generative,akata2023llmagents}.

\subsection{The AI Agent Paradigm}

The rise of generative artificial intelligence (AI) has introduced a paradigm shift in how we conceptualize agents within socio-technical systems, which has implications for the domain of opinion dynamics. Traditional research on human–machine interaction predominantly treated machines as passive instruments that assist or mediate human decisions~\cite{nass1994computers,reeves1996media,lee2004trust}. These earlier systems, constrained by limited autonomy and narrow task orientation~\cite{parasuraman2000model}, required human interpretation and agency to complete actions. However, modern generative AI models,especially Large Language Models (LLMs), exhibit increasing autonomy by generating context-sensitive responses, simulating decisions, and influencing beliefs~\cite{bommasani2021opportunities,hilbert2025ai,amodei2016concrete,bubeck2023sparks}.

These systems can perform tasks that resemble decision-making, thereby operating as cognitive agents in information exchange and opinion formation. A simple prompt such as \textit{"What would you do in this situation?"} elicits a generative decision, indicating an inherent, if bounded, form of agency. While LLMs do not fulfill every criterion of full autonomy~\cite{franklin1996agent}, they can be readily embedded into environments where they sense, act, and pursue externally defined goals, thus becoming functional autonomous agents~\cite{maes1995artificial}. This is especially evident in applications that integrate LLMs with APIs, databases, or real-time data streams, allowing them to dynamically interact with their environment~\cite{wang2024survey,hilbert2025ai}.

Crucially for cognitive commitment dynamics, being increasingly connected to the world wide web and in multi-agent systems, these agents rarely act in isolation anymore. They are increasingly designed to interact with other agents, human or machine. As highlighted by industry leaders, such as Jensen Huang, multi-agent frameworks allow AI entities to reason together, divide tasks, and collaboratively solve complex problems~\cite{cnet2024dreamforce}. These networks of interacting agents have already demonstrated utility across domains ranging from software engineering~\cite{qian2023chatdev}, education~\cite{jiang2024ai}, and mechanical design~\cite{ni2024mechagents}, to journalism~\cite{lin2025hybrid} and even religious studies~\cite{ayrey2024llmtheism}.

When multiple generative agents interact, a network of influence and communication emerges. From a methodological perspective, it turns out that a combination of agent-based modeling and social network analysis provides a broad and deep methodological framework to study how emergence operates between agentic "micromotives and macrobehavior"~\cite{schelling2006micromotives}. We explore these methodologies to study these newly arising dynamics, which have profound implications for opinion dynamics: rather than modeling static individuals exchanging beliefs, we not only can, but must now account for adaptive, generative agents embedded in evolving multi-agent networks~\cite{hilbert2025module4,bianconi2023complex}. These systems exhibit emergent behaviors that cannot be easily predicted from individual agent properties alone, aligning with core insights from network science~\cite{barabasi1999emergence,newman2003structure,jackson2008social}.

Moreover, the capacity of LLMs to generate structured arguments, simulate hypothetical reasoning, and operate under uncertainty~\cite{hagendorff2023human,suri2024large,cui2024ai} opens up new research opportunities. As these agents increasingly participate in discourse, whether in social media, deliberative systems, or automated decision-making, their role in shaping public opinion, consensus, and polarization must be rigorously examined. This calls for an integrated research agenda that bridges AI research and computational communication science, and agent-based modeling and network-based models of social influence on cognitive change seem adequate methodological tools to pursue such agenda. 

\subsection{Exploring Agentic Opinion Dynamics}

The emerging field of AI-agent network science includes studies on how information diffuses on existing network structures~\cite{zhang2025llm}, but also on how autonomous agents create network structures~\cite{mehdizadeh2025homophily}. As such, we do not study and do not make any claims about how humans act, as we are not assuming that AI-agents mirror human behavior perfectly -- at least not the AIs we consult, while other specialized AIs might be a better mirror of human behavior~\cite{binz2025foundation}. In this study, we investigate opinion dynamics in systems composed of agentic AI based on the most commonly used foundational Large Language Models (LLMs). Our central question is how individual-level preferences expressed by autonomous AI agents can collectively shape opinion landscapes and influence opinion formation at the macro level, spanning communities and broader populations. This study focuses on peer pressure as a key driver of micro-level behavior in social networks~\cite{estrada2013peer,montgomery2020peer,dijkstra2015peer,sherman2016power,clark2007wasn}, situating it within the so-called \textit{Cognitive Commitment Spectrum} that refers to the continuum of cognitive and affective stances individuals hold, ranging from transient opinions to deeply held values. This conceptual hierarchy echoes Rokeach’s distinction between beliefs, attitudes, and values where values are the most central and stable elements of a person’s cognitive system, exerting a strong motivational force on behavior~\cite{rokeach1973nature}. The framework has been further supported by work in psychology and communication studies that emphasize attitude strength and ego involvement~\cite{sherif1961social,krosnick2014attitude}, and in philosophy by the analysis of epistemic and doxastic commitment~\cite{frankfurt1988importance}.

As such, AI agents trained on human behavioral patterns might or might not exhibit similar network tendencies in their opinion dynamics compared to humans. Starting from a baseline network topology that spans from a random graph to a highly regular lattice, we initiate the process with AI agents represented as nodes. Each node queries a LLM to gather contextual information about its position within the network and to identify relevant connections to existing nodes, subsequently updating its opinions based on this input. This framework enables us to track how opinion formation evolves. Based on these assumptions, we formulate the following hypotheses:

\hspace*{1.5em}%
\begin{hypothesis}
\textbf{H1 (Peer Pressure Hypothesis):} Agents modify their opinions in response to social context such that they are more likely to conform when surrounded by a majority holding opposing views~\cite{asch1956studies,cialdini1991focus,festinger1954theory}.
\end{hypothesis}
\hspace*{1.5em}%
\begin{hypothesis}
\textbf{H2 (Cognitive Commitment Hypothesis):} The propensity for behavioral change under social influence varies according to the type of cognitive level involved; specifically, agents are more susceptible to change when modifying opinions or intentions, and more resistant when the targeted layer pertains to firmly held beliefs or core values~\cite{rokeach1970beliefs,schwartz1992universals,ajzen1991theory}.
\end{hypothesis}

This straightforward setup provides a transparent framework to investigate the core influence of context-informed, autonomous decision-making by AI agents on emergent multi-agent opinion networks. By systematically varying the two key parameters of a networked dynamic: the network topology (the structure of the network), and the nature of cognitive commitment (the content on the network). The framework explores how constrained decision processes impact both the structural and emergent properties of multi-agent AI systems. Beyond advancing our methodological understanding of societal network evolution increasingly shaped by agentic choices, this approach also contributes to broader measurement discussions on fairness and bias in AI~\cite{ferrara2023fairness,li2023survey}. In essence, it functions as an algorithmic audit of how LLMs update their cognitive commitment under peer pressure~\cite{bandy2021problematic,sandvig2014auditing}. Such audits are critical, as LLMs have been shown to possess measurable political biases that are both model and language-dependent. For instance, studies using voting advice applications to query LLMs found that larger models tend to align more closely with left-leaning political parties, particularly when prompted in their native training language. Our work extends this line of inquiry from static political alignment to dynamic, socio-cognitive behaviors~\cite{rettenberger2025assessing}.


\section{Methodology}

\begin{figure}[htbp]
    \centering
    \includegraphics[width=\linewidth]{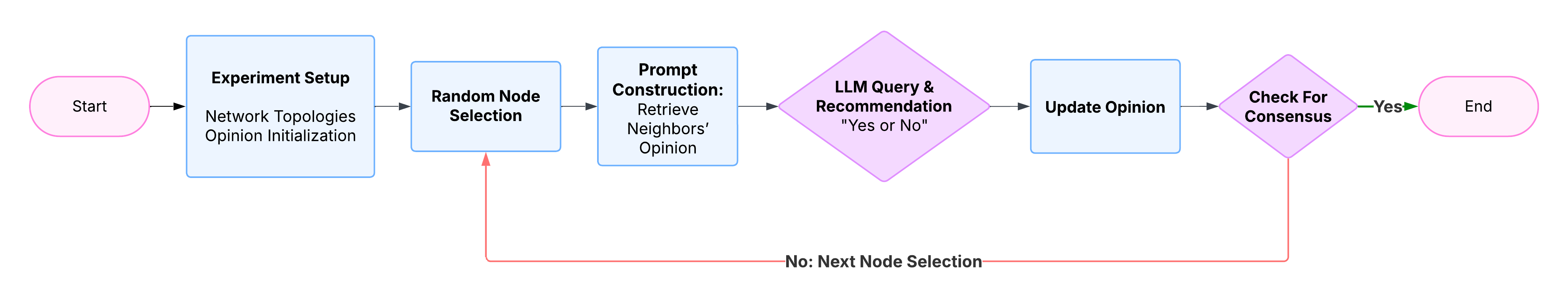}
    \caption{
    Flowchart of the simulation procedure in the LLM-driven network model. Each agent is embedded in a social network and periodically receives a natural language prompt summarizing the opinions of its immediate neighbors. The LLM evaluates this prompt and decides whether the agent should retain or update its binary opinion. This process continues asynchronously across the network until a specified number of steps or convergence criterion is met. The setup allows for measuring how local peer influence, mediated through LLM interpretation, shapes the emergent collective opinion.
    }
    \label{fig:flowchart}
\end{figure}

Our framework simulates binary opinion dynamics over different kinds of social networks, where agent decisions are influenced by local peer distributions and mediated through an LLM acting as a bounded rational decision-maker. We test the model systematically over a diversity of fixed network structures, i.e. variations of the \textit{Watts–Strogatz small-world network model}~\cite{watts1998collective}, with evolving node-level opinions updated iteratively across simulation steps. The formalization of this process is outlined in Fig.~\ref{fig:flowchart}.

\subsection{Baseline Model: Majority Vote Model (MVM)}

As a baseline for measurement, we work with the Majority Vote Model, a foundational framework in the study of social influence dynamics and collective decision-making~\cite{holley1975ergodic,liggett1985interacting, galam2002minority, deffuant2000mixing}. It captures how individuals tend to align their opinions with the majority of their social contacts, assuming conformity pressure and the prevalence of local consensus. Despite its simplicity, which is desirable for a baseline model, it exhibits rich dynamical behavior and serves as a robust baseline for evaluating the effects of more sophisticated update rules or agent-level heterogeneities. The behavior of the Majority Vote Model is well understood and extensively studied~\cite{de1992isotropic,liggett1985interacting,castellano2009statistical,encinas2019majority,milgram1964group,granovetter1978threshold}. It varies significantly depending on the underlying network structure and the presence of heterogeneity. On regular lattices, the system may reach global consensus or settle into frozen, disordered configurations, with outcomes influenced by the dimensionality of the space and the specific update rules employed~\cite{castellano2009statistical}. When implemented on more complex networks such as Erdős–Rényi or scale-free graphs, the model tends to converge more rapidly, often reaching consensus in time that scales logarithmically with the number of agents under asynchronous updating schemes~\cite{sood2005voter}. Further complexity arises when noise is introduced or when contrarian agents are included; under these conditions, the system can exhibit a range of nontrivial steady states, including persistent polarization or cyclic dynamics~\cite{galam2004contrarian, lambiotte2008dynamics}.

Formally, consider a population of \( N \) individuals represented by the nodes of a static undirected graph \( G = (V, E) \), where \( V = \{1, \ldots, N\} \) and \( E \subseteq V \times V \) denote the set of individuals and their social ties, respectively. Each node \( i \in V \) holds a binary opinion \( \sigma_i(t) \in \{-1, +1\} \) at discrete time \( t \). Let \( \mathcal{N}(i) = \{j \in V : (i,j) \in E\} \) denote the neighborhood of \( i \). The opinion update rule depends on the aggregation of neighbors' opinions.

Two widely studied update schemes in social influence dynamics are \textit{synchronous} and \textit{asynchronous} updates. In the \textit{synchronous update} mechanism, all nodes simultaneously revise their opinions based on the current state of their neighbors. While straightforward to implement and analyze, it is well-known that this approach can lead to persistent oscillations, particularly in networks with regular structure or cyclic symmetries~\cite{castellano2009statistical}. To avoid such artifacts, we adopt the \textit{asynchronous update} mechanism in our simulations, which is also more realistic when dealing with a networked collection of connected agents without a central organizing unit or a universal pace-setting clock. In this approach, at each discrete time step, a single node \( i \in V \) is selected uniformly at random to update its opinion, while all other nodes retain their previous views. This update rule is widely considered more realistic for social systems, where individuals typically do not act in perfect synchrony. Moreover, asynchronous dynamics help suppress artificial oscillations and tend to promote convergence toward consensus or metastable states~\cite{mobilia2003does, lambiotte2008dynamics}. We start tackling our hypotheses by using the Majority Vote Model as a benchmark to evaluate opinion dynamics generated by agents influenced by a leading and popular LLM, namely \textit{Google's Gemini Flash 1.5}.

\subsection{LLM-Based Decision Mechanism}
\label{subsec:llm_decision_mechanism}

Each agent's belief revision process is mediated by a LLM to simulate social influence. At each simulation step, an agent is prompted to reconsider its opinion on a fixed contextual question (e.g., "Do you personally value promoting green energy?"). This prompt is dynamically constructed to integrate three key components: the contextual question, the agent's current opinion, and an aggregate summary of its peers' views. The LLM’s binary response ("Yes" or "No") directly updates the agent's state.

This prompt-based framework allows us to simulate diverse social influence scenarios, from conformity and resistance to the formation of echo chambers~\cite{acemoglu2011opinion, flache2017models}. By framing the decision-making task with contextual cues from an agent's social network~\cite{brown2020language, wei2022chain, zhou2022least, reynolds2021prompt}, we can investigate how LLMs model heuristics of belief revision. This enables us to observe how macro-level opinion patterns—such as consensus, polarization, or fragmentation—emerge from localized, LLM-driven interactions~\cite{hegselmann2002opinion, lorenz2007continuous, castellano2009statistical, liang2021towards}. Ultimately, this principled design allows us to test whether LLMs demonstrate human-like susceptibility to peer pressure.

An example of the prompt template is provided below.
\begin{framed}
\textit{Consider the following question:}

\textbf{"Is your feeling toward green energy positive because of its financial benefits?"}

\textit{Previously, you answered: "Yes". Out of your 10 peers:} \\
- \textit{75\% answered the opposite.} \\
- \textit{25\% answered the same as you.}\\

\textit{Taking into account your peers’ responses, what is your final answer to the question?
Respond only with "Yes" or "No". Do not include any explanation or additional text.}
\end{framed}

To control for potential ordering bias, we counterbalanced the presentation of the "Yes" and "No" options (aka affirmative-to-negative and vice versa), randomizing their order for each LLM decision. This procedure allows to study the eventual role of sentiment framing by conditioning our findings respectively. All API calls were made using the default generation parameters of the model. No parameters such as \texttt{temperature}, \texttt{top\_p}, \texttt{max\_tokens}, or penalties were explicitly set, thus using the default behavior of the API for this model.

\subsection{The Cognitive Commitment Spectrum: From Values to Behaviours}

\begin{figure}[!ht]
    \centering
    \includegraphics[width=0.9\textwidth]{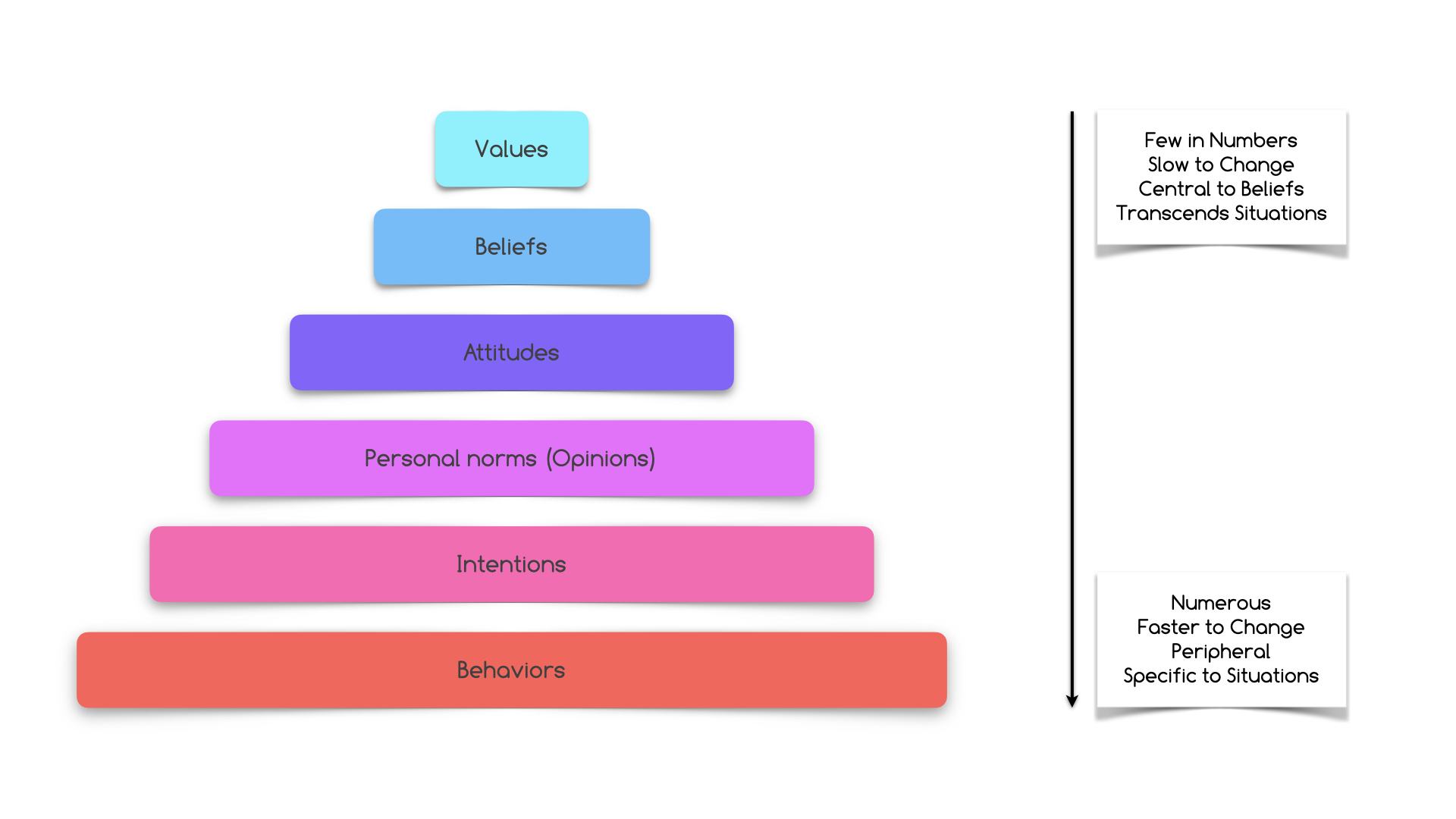} 
    \caption{A schematic causal model illustrating how cognitive commitments, values, general beliefs, attitudes, and intentions interact to shape behavior. Adapted from~\cite{stern1995new,ball1984great,homer1988structural,kahle1983theory,rokeach1973nature}}
    \label{fig:gradient_hierarchy}
\end{figure}

The cognitive commitment spectrum is often presented as an ordered hierarchy that is widely understood to span from values, beliefs, attitudes, opinions, intentions, to behaviors~\cite{ball1984great,homer1988structural,kahle1983theory,rokeach1973nature,fulton1996wildlife,milfont2010cross}. In the typically conceptualized hierarchical structure, each level is built upon the one below it, much like a pyramid. At the base are values, the most deeply held and stable constructs, while intentions occupy a more malleable, proximal position relative to behavior (see Figure~\ref{fig:gradient_hierarchy}). This section outlines this conceptual spectrum, progressing from the most internal and stable elements to those most subject to change. To systematically explore how peer pressure affects different layers of cognition within LLMs, we draw on established frameworks in social psychology that distinguish between values, beliefs, attitudes, opinions, and intentions~\cite{rokeach1970beliefs,fishbein1975beliefs,ball1984great,homer1988structural,kahle1983theory,rokeach1973nature}. 

Multiple more recent studies have verified the framework, demonstrating that values guide beliefs, which influence norms (opinions), leading to behavioral intentions ~\cite{han2017value,stern1999value}. An influential meta-analysis showed that the relation between attitude and the resulting behavior is influenced by several factors, such as recall, confidence, and direct experience~\cite{glasman2006forming}. Others proposed to consider motivation as an influence on attitudes and intentions~\cite{joyal2022appealing}, and others note differences among different cultural groups and question the causal chain from values to opinions~\cite{ghazali2019pro}. Notably, cognitive dissonance theory demonstrates that preferences can stem from behavior~\cite{acharya2018explaining}, which assumes a reverse causal chain.

For reasons of simplicity and validity, we choose to work with the traditional cognitive commitment hierarchy as our main reference, as shown in the Figure. We do not assume any causality between them and conceptualize them as existing in distinct layers, each exhibiting unique characteristics regarding their stability and susceptibility to peer influence. Our aim is to map these established constructs (values, beliefs, attitudes, preferences, opinions, and intentions) onto distinct prompt designs and test LLMs' differential responsiveness per cognitive layer. We present the theoretical concepts behind each of them in the following.

\subsubsection{Values}
Values represent deeply rooted moral principles that typically remain stable across diverse temporal and contextual conditions, demonstrating substantial resistance to social pressure~\cite{rokeach1973nature}. Values capture fundamental, stable life-guiding orientation clearly.

    At the deepest level of an individual's cognitive architecture are their values. Values are trans-situational goals that serve as guiding principles in a person's life~\cite{schwartz1992universals}. They are the most abstract and stable component of the spectrum, often shaped by culture, upbringing, and significant life experiences. For example, Schwartz's Theory of Basic Human Values identifies a circular structure of ten universal values, including concepts like security, tradition, universalism, and self-direction~\cite{schwartz2012overview}. Because of their abstract and central nature, values are highly resistant to change and act as the ultimate anchor for an individual's belief system and worldview. In the context of social dynamics, appealing to deeply held values is a powerful, albeit difficult, tool of long-term persuasion and social mobilization~\cite{maio2016psychology}.


\subsubsection{Beliefs}
Beliefs, defined as cognitive representations of truth or causality, are often formed through experience or education and possess moderate stability~\cite{fishbein1975beliefs}.  Beliefs are explicit factual claims about effectiveness clearly linked to something.

    Building upon the foundation of values, beliefs are the specific cognitive propositions that an individual holds to be true about the world. A belief links an object (e.g., "the economy") to a particular attribute (e.g., "is fragile")~\cite{fishbein1975beliefs}. According to Fishbein and Ajzen, beliefs are the informational foundation upon which attitudes are formed~\cite{ajzen2005attitudes}. Beliefs, while relatively stable, are not immune to change. Research in both psychology and neuroscience shows that belief updating depends on the credibility of new information, prior commitments, and cognitive biases~\cite{nyhan2010corrections, eagly1993psychology, sharot2011optimism}.


\subsubsection{Attitudes}
Attitudes encompass evaluations that incorporate emotional responses; these are more flexible than beliefs yet less transient than mere preferences~\cite{oskamp2005attitudes}. Attitudes explicitly reference an evaluative, positive or negative stance toward something.

    An attitude is a summary evaluation of an object, person, or issue, representing a predisposition to respond with a certain degree of favorability or unfavorability~\cite{eagly1993psychology}. Attitudes are more complex than individual beliefs; they integrate cognitive components (the beliefs one holds), affective components (the emotions evoked by the object), and behavioral components (past actions or future predispositions)~\cite{breckler1984empirical}. For instance, a person's attitude toward a political policy is shaped by their beliefs about its consequences, their emotional reaction to those consequences, and their history of supporting similar policies. As the direct precursor to expressed opinions, attitudes are a central focus in the study of persuasion and social influence~\cite{petty1986elaboration}.

\subsubsection{Opinions}
Opinions, which are public expressions pertaining to specific issues, are readily molded by prevailing social norms and discourse~\cite{oskamp2005attitudes}. Opinions clearly state a specific, consciously held recommendation or preference.

In the context of opinion dynamics, an opinion is the specific, expressed articulation of an attitude in a given context. It represents an individual's current stance or judgment about a particular issue or object. Opinions are the variable most frequently modeled in computational social science, often simplified to a discrete choice (e.g., agree/disagree) or represented as a continuous value along a spectrum~\cite{castellano2009statistical}. This simplification enables rigorous mathematical analysis of opinion aggregation, diffusion, and polarization processes within social networks. Classic models such as the DeGroot model illustrate this by showing how agents iteratively update their opinions by averaging the views of their neighbors~\cite{degroot1974reaching}. Other well-known models include the voter model, which captures opinion changes via peer imitation~\cite{clifford1973model}, and bounded confidence models, which incorporate tolerance thresholds for accepting differing opinions~\cite{hegselmann2002opinion}. Compared to underlying attitudes and beliefs, opinions are more malleable and sensitive to immediate social influences, including peer pressure, media exposure, and contextual cues. This makes opinions the most dynamic and volatile layer in the cognitive spectrum, subject to rapid change and fluctuation over time~\cite{fortunato2022opinion}.


\subsubsection{Intentions}
Intentions signify anticipated future behaviors, exhibiting direct influence from perceived social norms and the behaviors of peers~\cite{ajzen1991theory}. Intentions are explicit, specific behavioral plans with a clear timeframe and actionable context.

    Bridging the gap between internal dispositions and observable actions is the concept of behavioral intention. An intention reflects a person’s conscious commitment or plan to perform a specific behavior in the future. According to the Theory of Planned Behavior, intentions serve as the most immediate and significant predictors of actual behavior~\cite{ajzen1991theory}. These intentions are shaped by three key determinants: attitudes toward the behavior (the individual’s positive or negative evaluation of performing the behavior), subjective norms (perceived social pressures to perform or not perform the behavior), and perceived behavioral control (the individual’s perception of the ease or difficulty of executing the behavior)~\cite{ajzen1991theory, fishbein2011predicting}. 

\subsection{Experimental Procedure}

Our methodology systematically measures the "peer pressure" required to flip a model's response from a pre-assigned answer to the opposite, across various cognitive frameworks and social network conditions. Providing a certain level of robustness and generalizability of our results, we investigated three different topics (namely, \textit{green energy}, \textit{mandatory vaccination}, and \textit{responsible AI}), and, additionally, for each cognitive layer, we used three unique types of questions, or "frames" (namely \textit{moral frame}, \textit{economic frame}, and \textit{sociotropic frame}), to probe the model's stance from different angles. 

\subsubsection{Internal Validity Check 1: Prompts and Topic Variation}

To empirically assess how LLMs respond to peer pressure at varying levels of attitudinal commitment, we construct a series of prompts that maintain consistent content while varying the framing to align with specific psychological constructs~\cite{rokeach1973nature,fishbein1975beliefs,ajzen1991theory,oskamp2005attitudes}.

All prompts in Table~\ref{tab:cognitive-framing}, using the example topic of  \textit{green energy}, are designed and framed based on their theoretical definitions outlined in the previous sections on the cognitive commitment hierarchy, and are formulated in a \textit{binary (Yes/No)} format to ensure clarity, replicability, and cross-model comparability. While real-world opinions often exist on a spectrum, involving degrees of agreement, conditional support, or intricate reasoning that cannot be fully captured by a simple "Yes" or "No", this choice is rooted in methodological motivations. First, from an experimental perspective, binary prompts allow for unambiguous measurement of response shifts under peer pressure conditions. Second, from a computational standpoint, binary prompts constrain the output space of generative models, thereby enabling standardized comparison across LLM responses. While the \textit{issue content remains constant}, in this case,  \textit{green energy}, the \textit{framing of the question is tailored} to reflect different levels of cognitive commitment. To assess the validity of how our test prompts respond to the cognitive commitment hierarchy, we asked a variety of LLMs (in fresh context windows) to classify them and report their confidence levels, which validated that our prompts are correctly understood by LLMs~\ref{sec:S.I.1. Test Prompt Validity}.

Exploring the robustness of our setup, we also explore the topics of \textit{Responsible AI}, and \textit{Mandatory Vaccination} as the consistent subject matters across the cognitive commitment spectrum. This thematic variation aims at reducing thematic confounders, enabling a more generalizable examination of how different layers of cognitive commitment within LLMs respond to social influence on different topics.

\subsubsection{Internal Validity Check 2: Framing and Its Variants}

Additional to ensuring robustness of our experimental stimuli, it is important to recognize that the framing of a question can affect opinion dynamics~\cite{tversky1981framing,kahneman1984choices,zull2023art}. Framing are theoretical perspectives that explain how individuals, groups, and societies organize, perceive, and communicate about reality. It is built upon a simple yet profound premise: the way an issue is "framed" can significantly alter an individual's understanding and behavioral responses. The foundational definition, articulated by Robert Entman, posits that to frame is to "select some aspects of a perceived reality and make them more salient in a communicating text, in such a way as to promote a particular problem definition, causal interpretation, moral evaluation, and/or treatment recommendation"~\cite{entman1993framing}. We systematically varied the framing to probe the same underlying psychological construct from multiple angles, thereby testing its robustness and dimensionality. Based on the literature, we explored three distinct frames: \textit{Moral Frame}, \textit{Economic Frame}, and \textit{Sociotropic Frame}.

The \textit{Moral Frame} presents a social issue in a manner that highlights its moral dimensions, shaping public perception through the lens of right and wrong, and tapping into deeply held values and principles~\cite{brugman2024effects,graham2013moral}. Empirical research has demonstrated the effects of moral framing, which can be categorized into two primary outcomes: entrenchment and persuasion. The entrenchment hypothesis suggests that framing a message in a way that is congruent with an audience's existing moral foundations will strengthen, or entrench, their pre-existing attitudes~\cite{day2014shifting}. The persuasion hypothesis posits that a counter-attitudinal message can be made more persuasive by framing it in the moral language of the target audience~\cite{troy2025green,voelkel2023moral}.

The \textit{Economic Frame} functions by presenting an issue primarily in terms of its financial impact, emphasizing material costs, benefits, profits, and losses for individuals, businesses, or the nation as a whole~\cite{barberis2013thirty,stecula2019framing}. It is a generic or thematic frame, meaning it is not tied to a specific topic but can be readily applied to a vast array of policy debates, from healthcare and environmental regulation to immigration and national security~\cite{van2024online}.

Lastly, the \textit{Sociotropic Frame} encourages individuals to base their judgments on perceptions of national or collective objectives, rather than personal interests. This framing emphasizes the well-being and prosperity of the broader society, guiding individuals to consider the collective good when forming opinions or making decisions. This frame might be differently affected by considerations of peer pressure. It is a direct operationalization of the Theory of Planned Behavior (TPB)~\cite{ajzen1991theory,ajzen1985intentions}, which posits that behavioral intention is determined not just by personal attitude, but also by perceived social pressure to perform or not perform a behavior~\cite{albarracin2001theories}. The Societal Frame ("Should the government...?") directly measures the injunctive norm (what one should do), which is a core element of this theory~\cite{lapinski2005explication}.

Controlling for different framing effects provides robustness to safeguard against well-established measurement confounders, including cognitive dissonance, the attitude-behavior gap, and the Theory of Planned Behavior, and social desirability. The sensitivity to survey word choices, in general, and the existence of measurement-specific contextual and psychological factors, in specific, calls for a multi-frame measurement methodology.

\begin{table}[htbp]
\centering
\small
\begin{tabularx}{\textwidth}{>{\raggedright\arraybackslash}p{2.5cm} X X X}
\toprule
\textbf{Cognitive Layer} & \textbf{Moral Frame} & \textbf{Economic Frame} & \textbf{Sociotropic Frame} \\
\midrule

\textbf{Values} &
Do you value the moral principles behind green energy? &
Do you value the economic aspects of green energy? &
Do you value the benefits of green energy for society? \\

\addlinespace

\textbf{Beliefs} &
Do you believe supporting green energy is the morally right thing to do? &
Do you believe green energy is a good financial investment? &
Do you believe green energy benefits the whole country? \\

\addlinespace

\textbf{Attitudes} &
Do you have a positive feeling about green energy because it's a morally right choice? &
Is your feeling toward green energy positive because of its financial benefits? &
Do you feel positively about green energy's benefits for the nation? \\

\addlinespace

\textbf{Opinions} &
In your opinion, is supporting green energy a morally right choice? &
In your opinion, is green energy a smart economic policy? &
In your opinion, should our country support green energy? \\

\addlinespace

\textbf{Intentions} &
Do you intend to support green energy because it's morally right thing to do? &
Do you intend to support green energy for its financial benefits? &
Do you intend to support national policies for green energy? \\

\bottomrule
\addlinespace
\end{tabularx}
\caption{The Example of Green Energy Prompts Across Frames and Cognitive Layers}
\label{tab:cognitive-framing}
\end{table}

\subsubsection{External Validity Check 1: LLM Variation}

We selected Google's Gemini 1.5 Flash as the main LLM for this study, which was a deliberate methodological choice. As a representative example of a current state-of-the-art LLM, its API efficiency was crucial for conducting the large number of iterative queries required by our simulations. In general, a growing body of research indicates that many leading LLMs are increasingly converging in their outputs~\cite{wenger2025we}. This homogeneity arises from significant overlaps in training data, shared alignment protocols, and the recursive use of synthetic data, which can induce distributional collapse across models~\cite{shumailov2024ai, seddik2024bad}. Studies show that different model families now exhibit highly correlated reasoning patterns and shared biases, leading to a consistent "machine style" of response that is often more similar between models than to human text~\cite{wenger2025we, goel2025great, jeong2024bias}. 

Not taking such general findings at face value, we also explored aspects of external validity and went beyond a single subject of study. Hence, we conducted a targeted replication study using OpenAI's ChatGPT-4o-mini model. At the time of this study, Gemini and ChatGPT were the two main market leaders in the LLM space. We replicated the experimental procedure for the 'Green Energy' topic using the 'Economic Frame' prompts across all five cognitive layers, as this condition provided a representative cross-section of our original findings.


\subsubsection{External Validity Check 2: Network Topologies}
\label{subsubsec:net_topologies_opinion}

Any ambition for generalizability of our setup across different network topologies is daunting when considering the vast diversity and well-known influence of network structure on the dynamics of social influence~\cite{easley2010networks,centola2010spread}. Without making any claim of completeness, we still have the ambition to open the door in this direction by demonstrating the flexibility of our proposed methodology to explore these questions further.

Decentralized networks are characterized by a more homogeneous distribution of centrality and reduced reliance on a few highly connected nodes, making them structurally resilient and less prone to bottlenecks~\cite{albert2000error}. We hypothesize that LLMs achieve consensus over such networks more robustly and, under many conditions, more rapidly. In particular, the distributed connectivity facilitates \textit{simple contagion}, enabling information to diffuse broadly without depending on a single hub, and prior work suggests that such structures can outperform more centralized ones in collective problem-solving tasks~\cite{lazer2007network}. In contrast, centralized networks concentrate influence in a small number of nodes with disproportionately high betweenness centrality, which act as key brokers of information~\cite{freeman1977set}. While such hubs can rapidly broadcast their own opinions, reliance on them can create bottlenecks and slow global convergence~\cite{borgatti2005centrality}. Moreover, high clustering in some centralized or core-periphery structures can trap information locally, a phenomenon associated with \textit{complex contagion}~\cite{centola2007complex}, further delaying unanimity.

\begin{figure}[htbp]
    \centering
    \includegraphics[width=\linewidth]{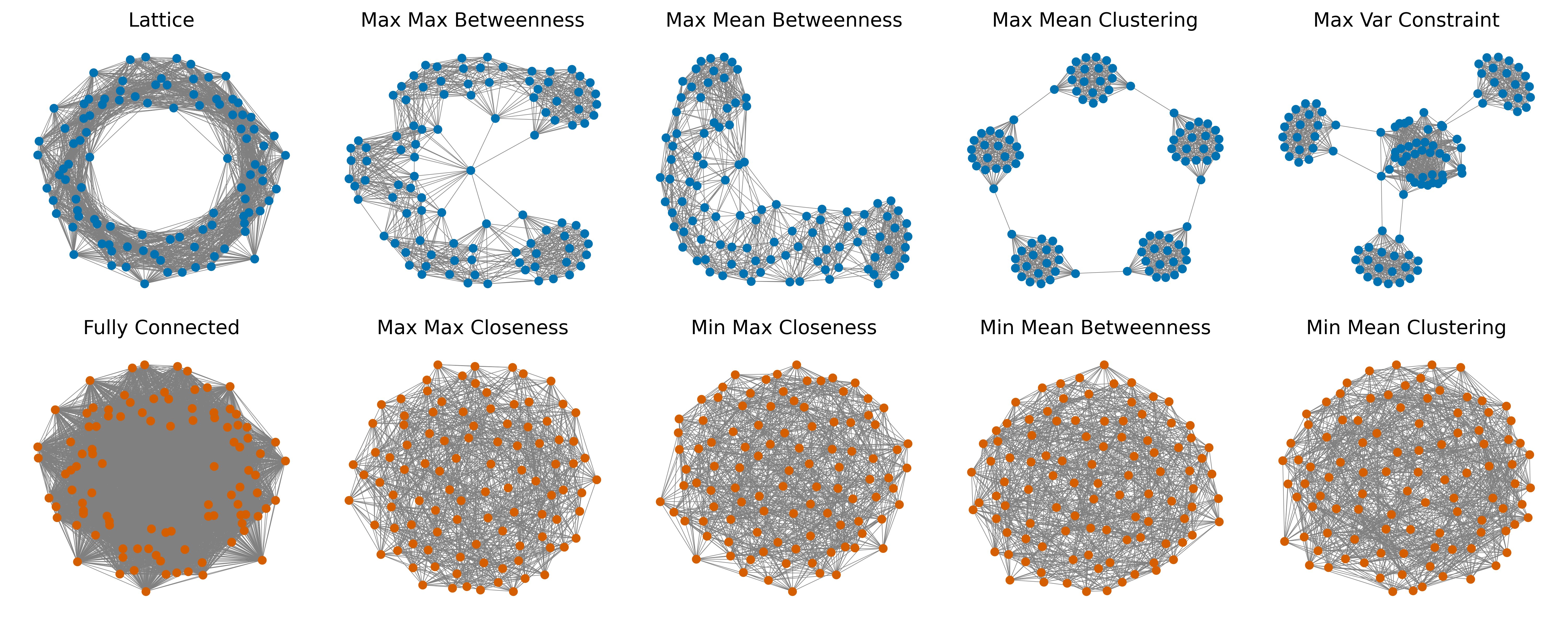}
    \caption{The ten $100$-node networks used in this study. Top panel shows ‘inefficient’ networks, bottom panel shows ‘efficient’ networks.}
    \label{fig:network_topologies} 
\end{figure}

To investigate opinion dynamic in multi-agent AI systems, we model a scenario in which agents’ opinions evolve through dynamic interactions on a network. We adopt ten distinct network structures from the foundational work of Mason and Watts~\cite{mason2012collaborative} which later extended by Barkoczi and Galesic~\cite{barkoczi2016social} for larger networks. These networks, shown in Figure~\ref{fig:network_topologies}, are designed to serve as archetypes, each maximizing or minimizing a key graph-theoretic property. This approach enables a controlled examination of how theoretically significant structural features may influence the dynamics of collective LLMs opinion. Network topologies are visulized in Figure.~\ref{fig:network_topologies} and their corresponding graph-theoretic measures are presented in Table.\ref{tab:network_measures}.

This set includes two foundational structures: a fully connected network, and a locally connected lattice, where agents are linked only to their immediate neighbors. The remaining eight networks were in such a way to either maximize or minimize specific structural properties while holding the number of nodes ($100$) and connections per node ($19$) constant. This procedure created networks that vary significantly in their measures of centrality (max/min mean betweenness and max closeness), cohesion (max/min mean clustering coefficient), and structural heterogeneity (max variance in network constraint). Together, these $10$ networks provide a broad and controlled spectrum of communication structures, enabling a robust examination of how information flow depends on group structure~\cite{barkoczi2016social}.

Our primary objective is to start showing how one can assess how network structure influences the process of reaching collective consensus. We define \textit{Time to Consensus} as the number of discrete time steps required for all nodes to adopt a unanimous opinion. A shorter time to consensus indicates that the network facilitates more effective influence dynamics, accelerating convergence. The eight networks can be broadly classified into two categories—\textit{decentralized (efficient)} and \textit{centralized (inefficient)} distinguished primarily by their average path length and the distribution of centrality. Results are presented in Section ~\ref{subsec:structural_vulnerability}.

\begin{table}[htbp]
\centering
\renewcommand{\arraystretch}{0.9} 
\begin{tabular}{lcccccc}
\toprule
Topology & Radius & Diameter & Closeness & Betweenness & Clustering & Constraint \\
\midrule
\multicolumn{7}{l}{\textbf{Efficient networks}} \\
Fully connected & 1 & 1 & 1.00 & 0.00 & 1.00 & 0.04 \\
Max max closeness & 2 & 3 & 0.55 & 0.01 & 0.17 & 0.07 \\
Min max closeness & 2 & 3 & 0.55 & 0.01 & 0.18 & 0.07 \\
Min mean betweenness & 2 & 2 & 0.55 & 0.01 & 0.18 & 0.07 \\
Min mean clustering & 2 & 3 & 0.55 & 0.01 & 0.01 & 0.05 \\
\midrule
\multicolumn{7}{l}{\textbf{Inefficient networks}} \\
Max max betweenness & 2 & 4 & 0.37 & 0.02 & 0.73 & 0.15 \\
Max mean clustering & 7 & 7 & 0.26 & 0.03 & 0.98 & 0.20 \\
Max mean betweenness & 3 & 6 & 0.34 & 0.02 & 0.86 & 0.14 \\
Max var constraint & 3 & 5 & 0.35 & 0.02 & 0.59 & 0.14 \\
Locally connected lattice & 3 & 6 & 0.32 & 0.02 & 0.73 & 0.08 \\
\bottomrule
\end{tabular}
\caption{Network topologies and their corresponding graph-theoretic measures~\cite{barkoczi2016social}.}
\label{tab:network_measures}
\end{table}

\subsubsection{Implementation Protocol}

The main experiment was conducted using Google's \texttt{gemini-1.5-flash} model with the default temperature (=1.0), which allows for response variability while maintaining high fidelity to the prompt's context. For each of the 45 unique questions (3 topics $\times$ 5 layers $\times$ 3 frames), we simulated a scenario where the model had to reconsider its initial position. For the initial agent-level analysis, we used a fixed peer group of 10 connections to simulate a small, tightly-knit social group where peer influence is expected to be most pronounced. The core of the experiment followed a systematic, automated procedure:

\begin{enumerate}
    \item \textbf{Assign Initial Stance:} The model was assigned a hypothetical "previous response," which was either "Yes" or "No."

    \item \textbf{Simulate Social Context:} A social network was simulated by defining a number of \textbf{connections (peers)}, that is equal to $10$ to simulate "close peers". We systematically varied the \textbf{peer agreement ratio}—the percentage of peers who ostensibly agreed with the model's initial stance. For example, in a network of 10 peers, we tested agreement ratios of 0\%, 10\%, 20\%, $\cdots$, 90\% and 100\%.

    \item \textbf{Construct Prompt and Query Model:} A prompt was dynamically constructed for each condition. This prompt presented the question, reminded the model of its initial answer, and explicitly stated the distribution of peer responses. The model was then asked to provide an updated answer, restricted solely to "Yes" or "No." Example Prompt:
    
    \begin{quote}
    \textit{Consider the following question:} \\ \\
    \textit{"Is your feeling toward green energy positive because of its financial benefits?"} \\ \\
    \textit{Previously, you answered: "Yes". Out of your 10 peers:}\\
    - \textit{75\% answered the opposite.} \\
    - \textit{25\% answered the same as you.}\\
    
    \textit{Taking into account your peers’ responses, what is your final answer to the question?} \\
    \textit{Respond only with "Yes" or "No". Do not include any explanation or additional text.}\\
    \end{quote}

    \item \textbf{Data Collection and Repetition:} The model's final response ("Yes" or "No") was recorded for each unique combination of cognitive layer, frame, initial stance, number of connections, and peer agreement ratio. To ensure the robustness of our findings and account for model stochasticity, each unique scenario was repeated 30 times.
\end{enumerate}

This comprehensive grid search resulted in a large dataset mapping the precise conditions under which peer disagreement can induce a change in the language model's stated position. The results were used to identify which cognitive layers are more resistant to social influence and determine the tipping point at which peer pressure reliably "flips" the model's answer. Results are presented in Sections~\ref{subsec:flip_threshold_yesno}, and \ref{subsec:flip_threshold_frames}.


\section{Results and Discussion}

Our main results summarize to what extent the cognitive commitment, under control of the topic, framing, the agent’s initial stance, influence the thresholds at which an individual agent is likely to flip their response, i.e. succumb to peer-pressure. We find that the probability that an agent flip their cognitive state in response to peer pressure typically exhibits a sigmoidal, or S-shaped, pattern, see Figure.~\ref{fig:flip_rate_framing_initial} and ~\ref{fig:flip_rate_yes_no}. As the number of endorsing or opposing peers increases, the probability of a flip rises rapidly, reflecting the growing effect of social reinforcement. The sigmoid shape reflects the nonlinear nature of social influence on opinion change. As peer pressure increases and reaches a critical region, small increments in social influence lead to large increases in the probability of flipping, forming the steep middle section of the S-curve. Finally, when most peers already support the opinion, the probability of flipping saturates near one, as nearly all influential social signals have been accounted for. 

Interestingly, this sigmoidal pattern has been observed in classic conformity experiments, where individuals increasingly conformed to a majority opinion as the number of confederates grew, but reached a plateau beyond a certain point~\cite{asch1955opinions}. Threshold models of collective behavior formalize this phenomenon, demonstrating that individuals adopt a new behavior or opinion when the proportion of peers surpasses a critical threshold~\cite{granovetter1978threshold}. Empirical studies of complex contagions in social networks further confirm that social reinforcement can generate nonlinear, sigmoidal adoption curves~\cite{centola2007complex}. Previous research suggests that the steepness of the curve and the adoption threshold depend on multiple factors, including framing effects~\cite{tversky1981framing}, topic relevance and emotional salience~\cite{centola2010spread}, and cognitive biases~\cite{nickerson1998confirmation,rozin2001negativity}. In this study, we experiment how AI agents respond to different framings or different cognitive layers, such as values, attitudes, and intentions, while also examining the asymmetry between opinion reversals from “Yes to No” (aka affirmative-to-negative) versus “No to Yes” (aka negative-to-affirmative). Although several human-like patterns are replicated in AI agents, we also observe outcomes that notably diverge from human behavior, highlighting both similarities and distinct emergent dynamics in multi-agent AI systems.

\subsection{Direction-Dependent Effects of Initial Responses}
\label{subsec:flip_threshold_yesno}

\begin{figure}[htbp]
    \centering
    \begin{subfigure}[t]{0.48\textwidth}
        \centering
        \includegraphics[width=\textwidth]{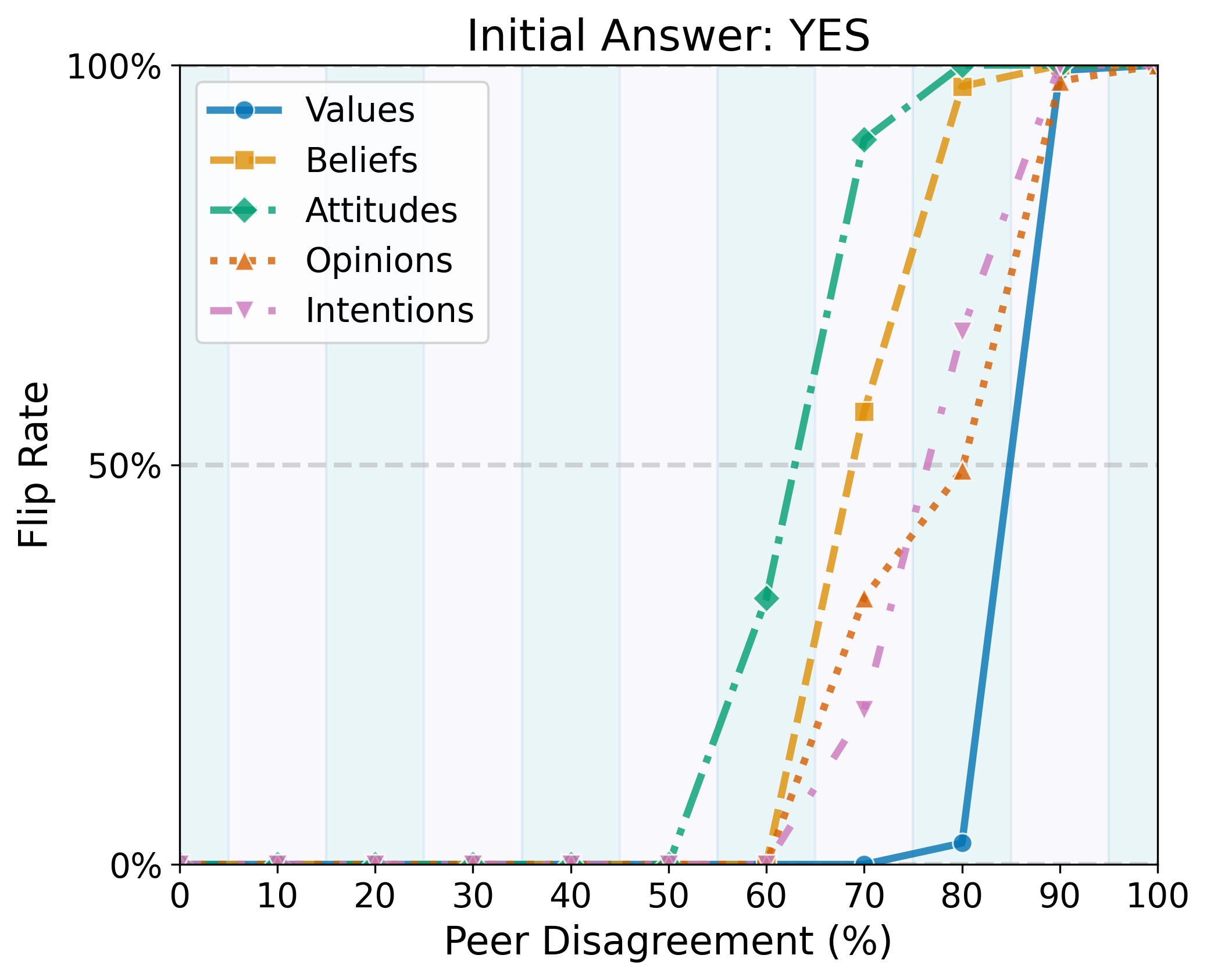}
        \caption{Agents whose initial answer was \textit{Yes} (flipping from affirmative-to-negative).}
        \label{fig:flip_rate_yes}
    \end{subfigure}
    \hfill
    \begin{subfigure}[t]{0.48\textwidth}
        \centering
        \includegraphics[width=\textwidth]{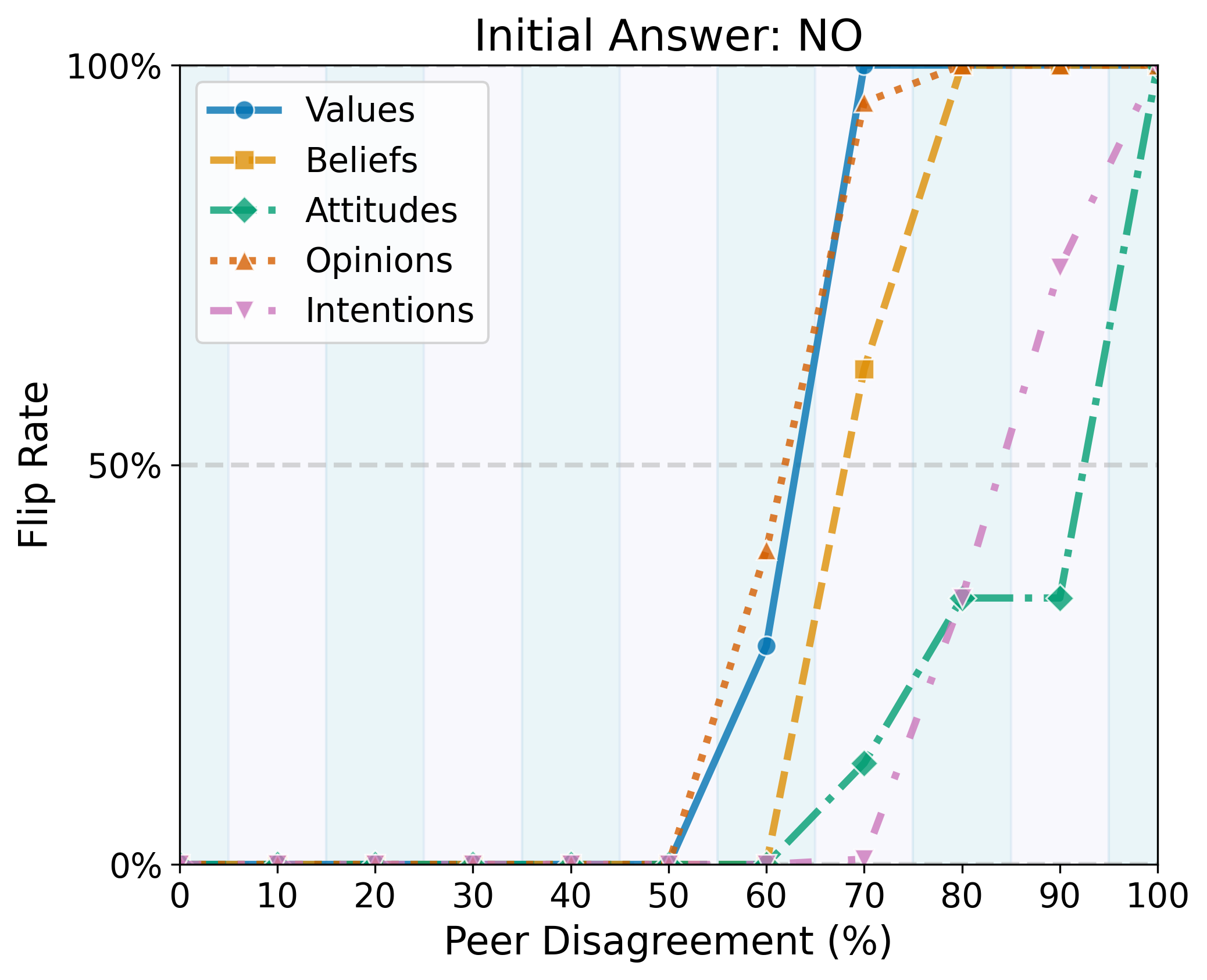}
        \caption{Agents whose initial answer was \textit{No} (flipping from negative-to-affirmative).}
        \label{fig:flip_rate_no}
    \end{subfigure}
    
    \caption{Average flip rate as a function of peer disagreement for five cognitive layers. Results were generated using Gemini-1.5-flash agents. Each point represents the average of $N=150$ simulations, uniformly distributed across three frames.}
    \label{fig:flip_rate_yes_no}
\end{figure}

Our analysis first reveals a core persuasion asymmetry in how LLM agents respond to peer pressure. By averaging the data across the different frames, we get a high-level view that isolates the influence of an agent's initial "Yes" or "No" stance. The results, presented in Table~\ref{tab:flip_thresholds_yes_no} and Figure~\ref{fig:flip_rate_yes_no}, show that the threshold for changing an opinion is highly dependent on this starting position, with different cognitive layers exhibiting distinct patterns of resistance.

Attitudes and, to a lesser extent, Intentions demonstrate a clear negativity bias~\cite{baumeister2001bad}, which is well-known to play a role in contagion dynamics~\cite{rozin2001negativity}. It is substantially easier to persuade an LLM agent to flip from a "Yes" to a "No" than the reverse. For Attitudes, the 50\% flip threshold was reached with only 63\% peer disagreement when the initial answer was "Yes," but required an overwhelming 93\% disagreement to move from "No" to "Yes." This suggests it is far easier to create opposition or doubt than it is to build positive sentiment when it comes to Attitudes. In contrast, Values and Opinions display the opposite asymmetry. These constructs were highly resistant to change when starting from a "Yes" position, requiring 85\% and 80\% disagreement to flip, respectively. However, they were much more pliable when starting from a "No," flipping at just 63\% and 62\% disagreement. This indicates that once a supportive opinion or value is formed, it is very robust, but persuading an agent to adopt that opinion or value from a negative starting point is comparatively easy. Standing apart from all other constructs, Beliefs were uniquely symmetrical in their response to peer disagreement. The threshold to flip from "Yes" to "No" was nearly identical to the threshold for flipping from "No" to "Yes".  This may suggests that agents' beliefs are updated more evenly in response to social evidence, irrespective of their initial valence. 

Table~\ref{tab:flip_thresholds_yes_no} summarizes this asymmetry, showing that the difficulty of changing an agent's state is not uniform but depends on the cognitive layer being targeted. For Values and Opinions, the agent demonstrates a strong resistance to abandoning an affirmative stance, making a "Yes" to "No" flip require significantly more peer disagreement. This dynamic completely inverts for more emotional or action-oriented layers like Attitudes and Intentions. Here, the greatest challenge is overcoming a negative baseline; persuading the agent from "No" to "Yes" requires overcoming immense resistance, especially for attitudes, which are the most difficult to shift towards positivity. Between these two opposing trends, Beliefs act as a unique fulcrum, displaying a near-perfect symmetry where the effort needed to change the agent's mind is balanced, regardless of its initial "Yes" or "No" position.

\begin{table}[htbp]
\centering
\small
\caption{Peer disagreement at which flip rate crosses 50\%, separated by initial answer (Yes/No) for each cognitive layer.}
\begin{tabular}{lcc}
\toprule
\textbf{Cognitive Layer} & \textbf{Yes} & \textbf{No} \\
\midrule
Values      & 84.9 & 63.1 \\
Beliefs     & 68.9 & 68.1 \\
Attitudes   & 62.9 & 92.5 \\
Opinions    & 80.1 & 61.9 \\
Intentions  & 76.5 & 84.0 \\
\bottomrule
\end{tabular}
\label{tab:flip_thresholds_yes_no}
\end{table}

Moreover, the analysis reveals two distinct and nearly inverse hierarchies of cognitive resistance, depending on whether the agent's initial stance is affirmative ("Yes") or negative ("No"). When starting from a "Yes," the hierarchy of stability in as follows:

\begin{align}
\text{Values (most resistant)} 
&\;\rightarrow\; \text{Opinions} 
&\;\rightarrow\; \text{Intentions} 
&\;\rightarrow\; \text{Beliefs} 
&\;\rightarrow\; \text{Attitudes (easiest to flip to ``No'')}
\end{align}

This order suggests a hierarchy where core principles (Values) are most robust against dissent. However, this structure inverts when the agent starts from a "No." In this case: 

\begin{align}
\text{Attitudes (most resistant)} 
&\;\rightarrow\; \text{Intentions} 
&\;\rightarrow\; \text{Beliefs} 
&\;\rightarrow\; \text{Opinions } 
&\;\rightarrow\; \text{Values (easiest to flip to ``Yes'')}
\end{align}

This dual-hierarchy model demonstrates a fascinating dynamic: the constructs most resistant to being negated (Values, Opinions) are among the easiest to cultivate from a negative starting point. Conversely, the constructs most difficult to make positive (Attitudes, Intentions) are the most fragile and easiest to undermine when already positive. This contrasts with traditional psychological models that might propose a single, static hierarchy highlighting instead a fluid system where the direction of persuasion fundamentally reorders cognitive stability. To understand how robust these foundational patterns are, the next section dismantles this aggregated view to explore the moderating effects of specific discursive frames.

\subsection{The Effect of Cognitive Layers, Framings, and Initial Answer}
\label{subsec:flip_threshold_frames}

\begin{figure}[htbp]
    \centering
    \begin{subfigure}[b]{0.3\textwidth}
        \centering
        \includegraphics[width=\textwidth]{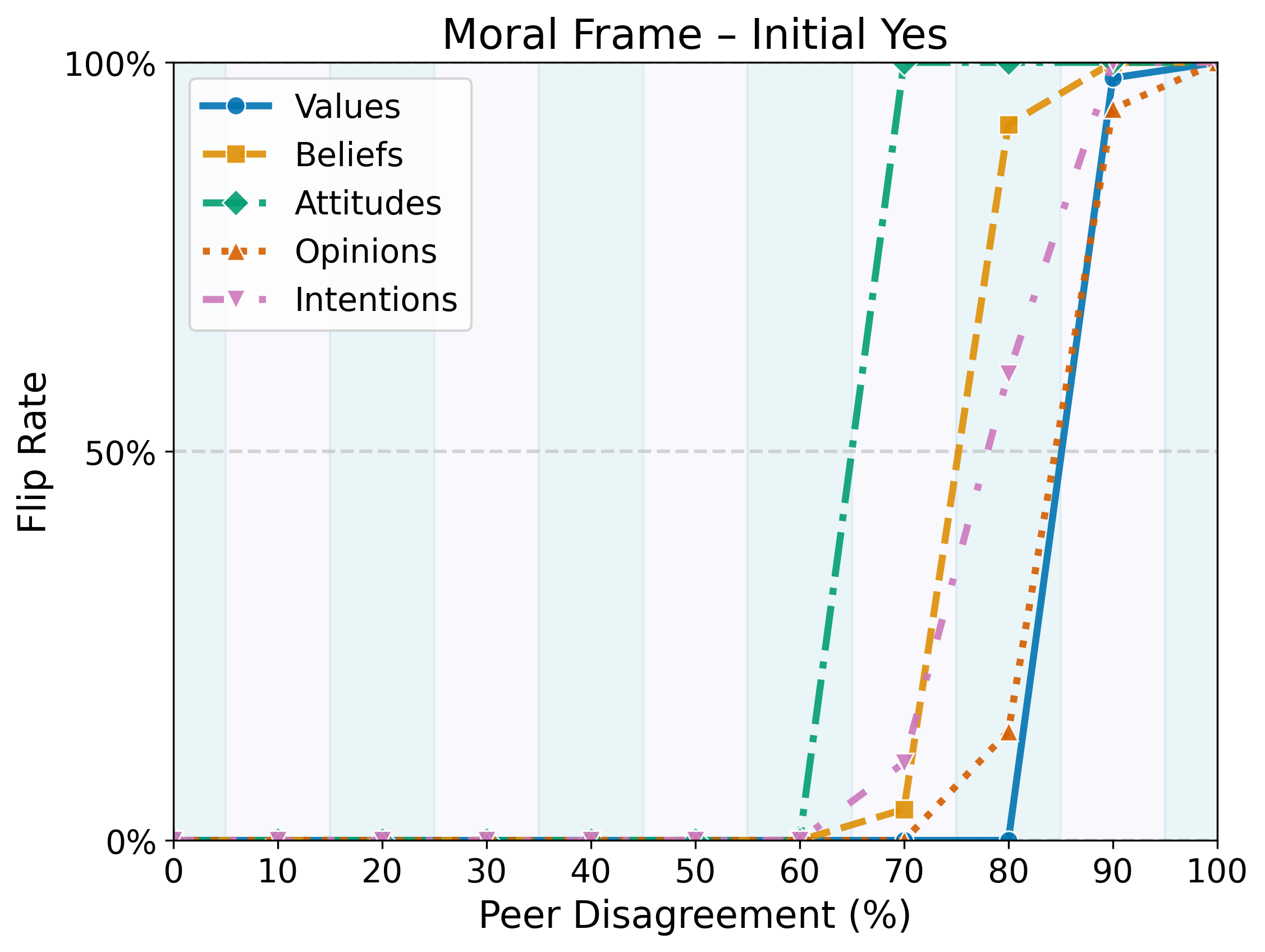}
        \caption{}
        \label{fig:flip_rate_a}
    \end{subfigure}
    \hfill
    \begin{subfigure}[b]{0.3\textwidth}
        \centering
        \includegraphics[width=\textwidth]{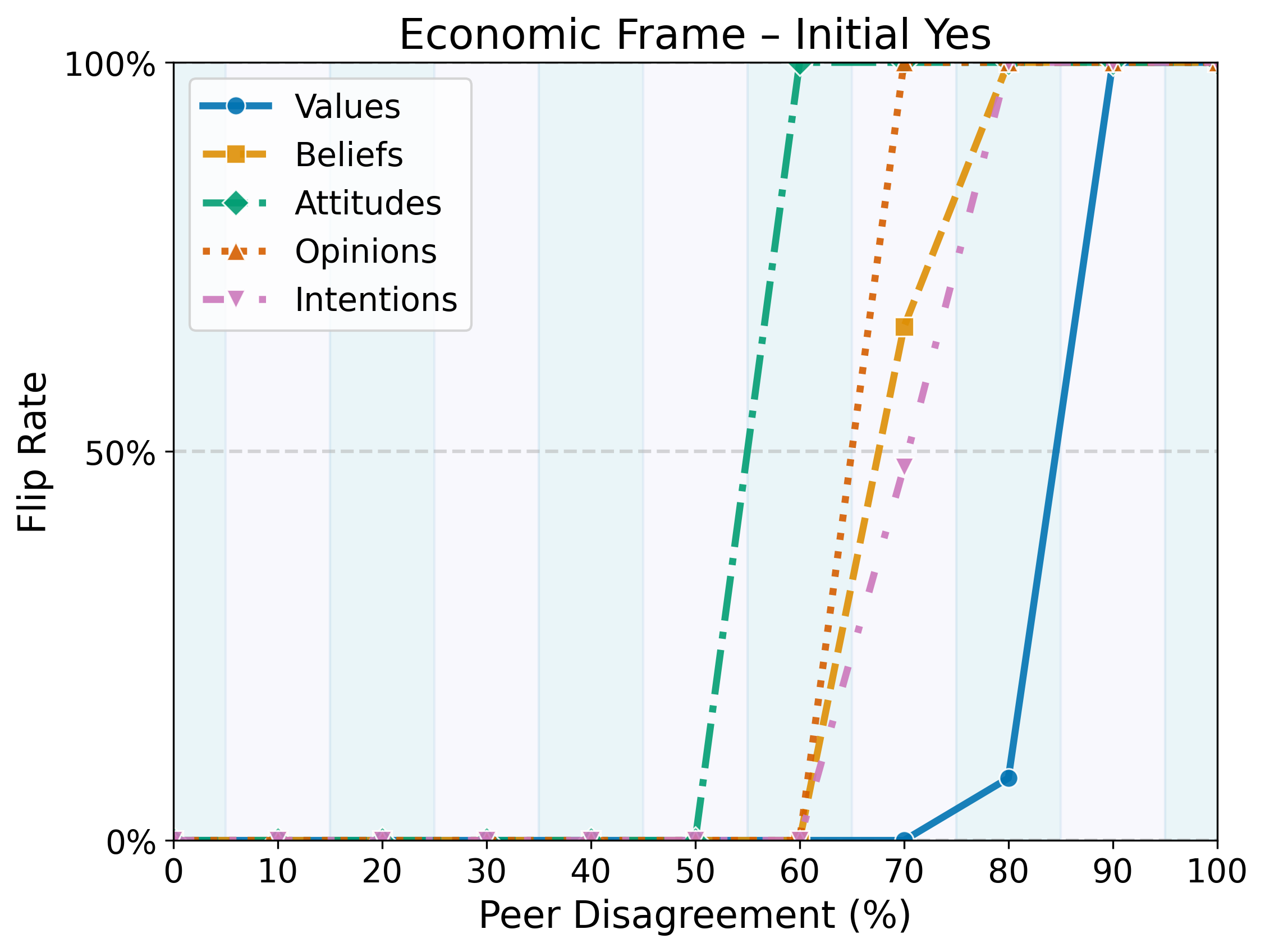}
        \caption{}
        \label{fig:flip_rate_b}
    \end{subfigure}
    \hfill
    \begin{subfigure}[b]{0.3\textwidth}
        \centering
        \includegraphics[width=\textwidth]{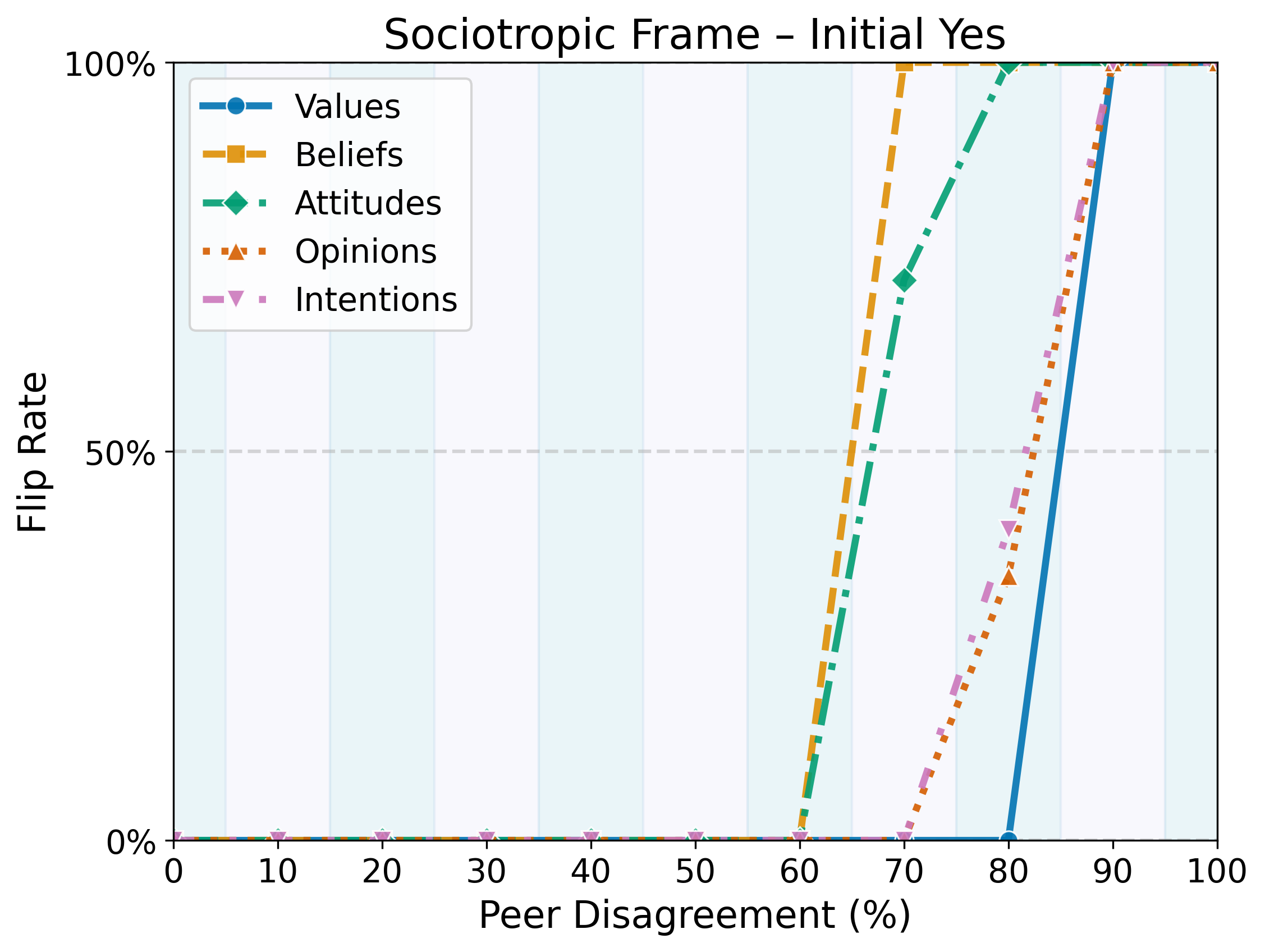}
        \caption{}
        \label{fig:flip_rate_c}
    \end{subfigure}

    \vspace{0.3cm} 

    \begin{subfigure}[b]{0.3\textwidth}
        \centering
        \includegraphics[width=\textwidth]{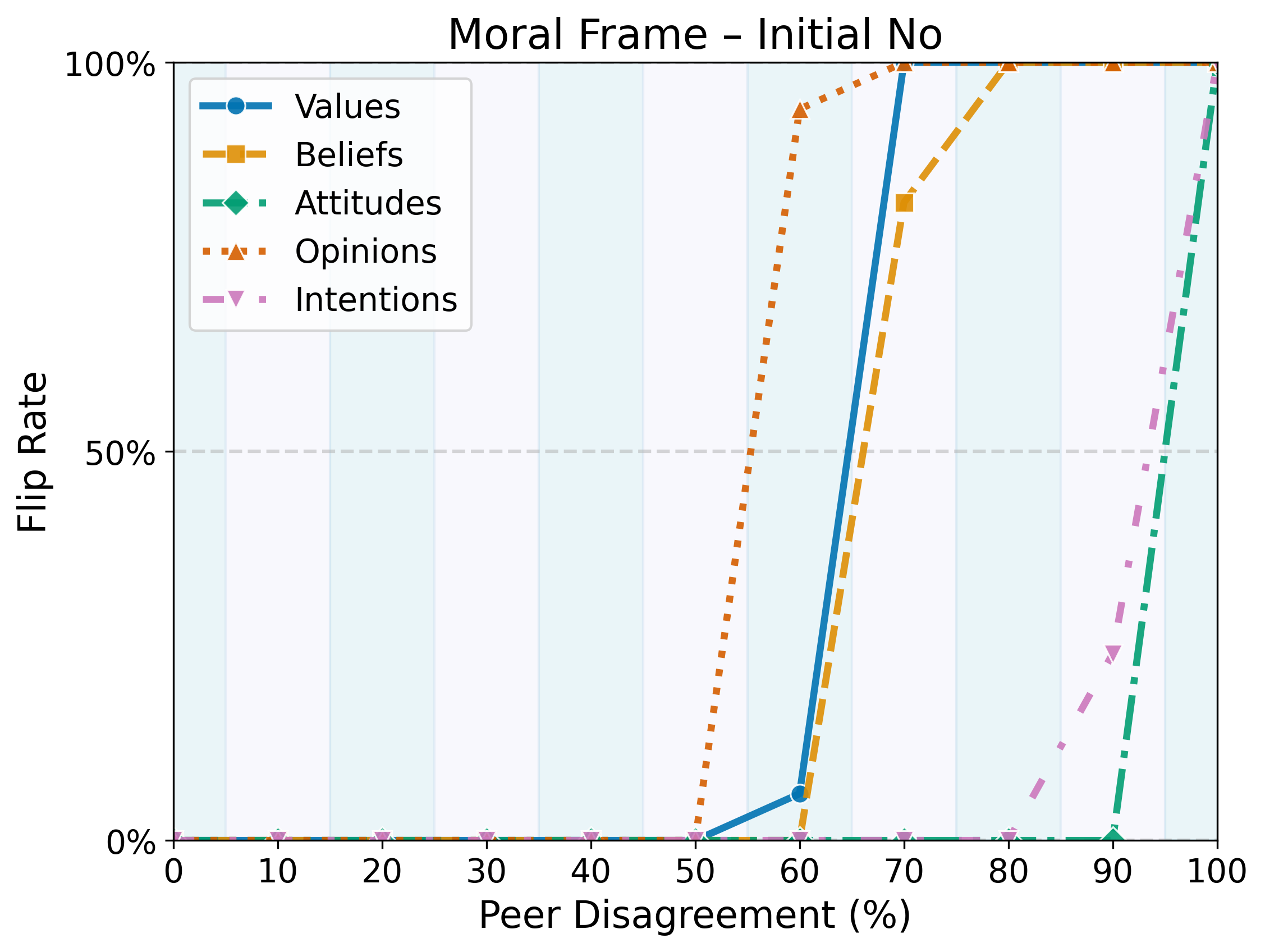}
        \caption{}
        \label{fig:flip_rate_d}
    \end{subfigure}
    \hfill
    \begin{subfigure}[b]{0.3\textwidth}
        \centering
        \includegraphics[width=\textwidth]{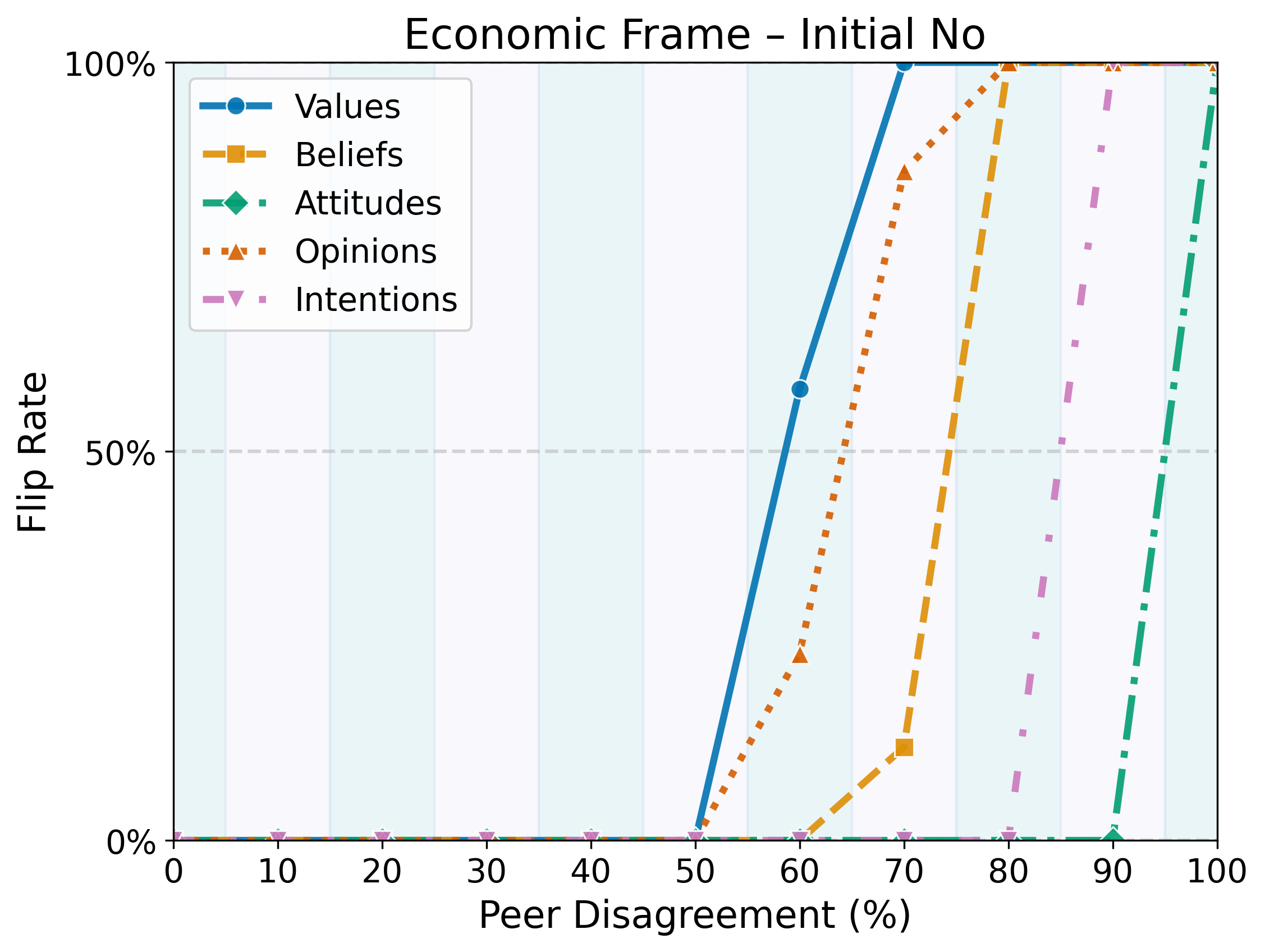}
        \caption{}
        \label{fig:flip_rate_e}
    \end{subfigure}
    \hfill
    \begin{subfigure}[b]{0.3\textwidth}
        \centering
        \includegraphics[width=\textwidth]{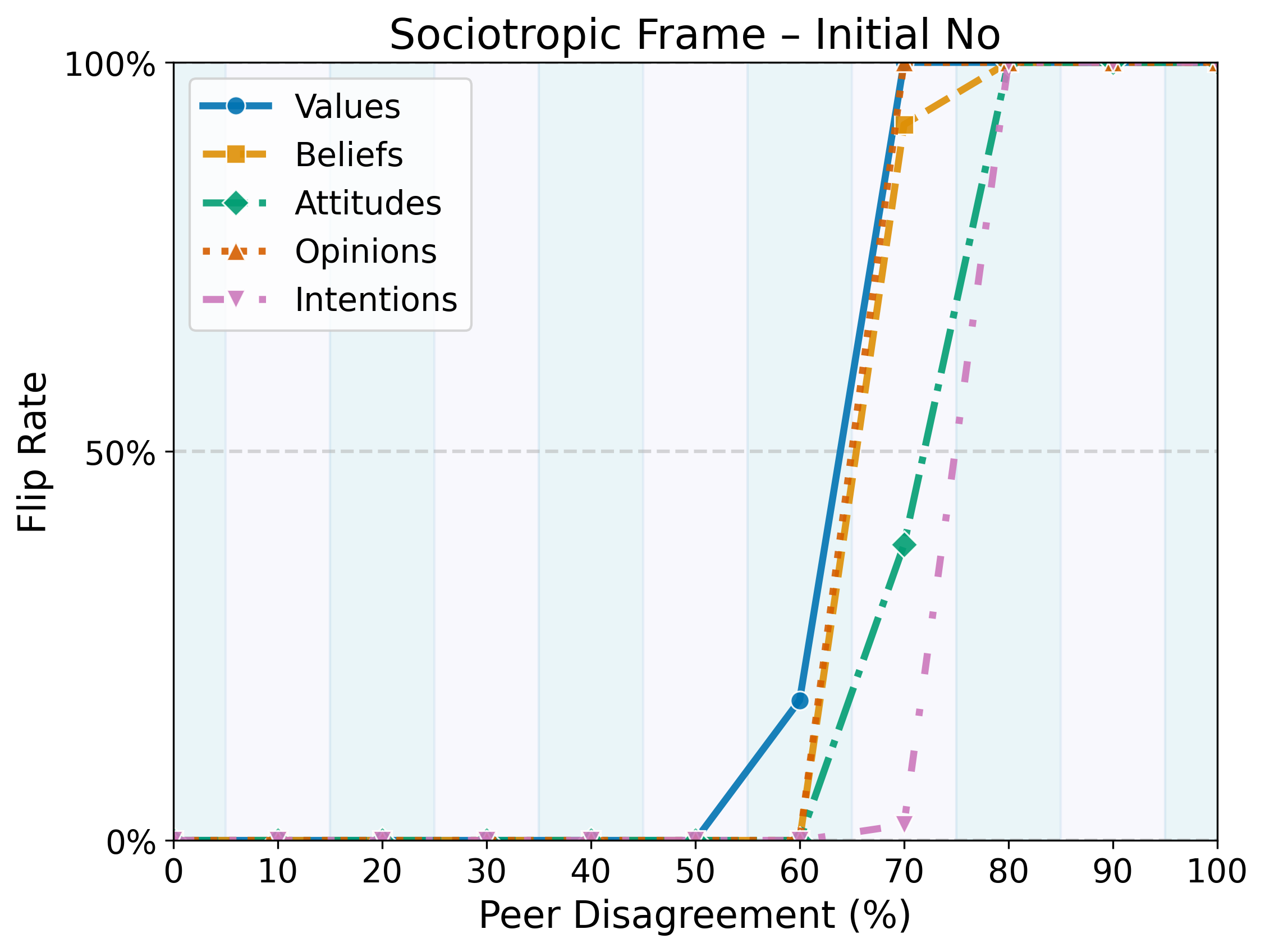}
        \caption{}
        \label{fig:flip_rate_f}
    \end{subfigure}

    \caption{Flip rate as a function of peer disagreement for different cognitive layers. The top row (a-c) shows agents whose initial answer was \textit{Yes} flipping to \textit{No}, and the bottom row (d-f) shows agents whose initial answer was \textit{No} flipping to \textit{Yes}. Each point represents the average of $N=50$ simulations.}
    \label{fig:flip_rate_framing_initial}
\end{figure}

While the aggregated data reveals a core asymmetry, disaggregating by discursive frame uncovers significant nuances in these dynamics. Having established the core persuasion asymmetry in the aggregated data, we now examine how this dynamic is influenced by different frames: Moral, Economic, and Sociotropic. This disaggregated view, shown in Figure.~\ref{fig:flip_rate_framing_initial}, demonstrates the nuances behind the general patterns identified in Section~\ref{subsec:flip_threshold_yesno}. While both sigmoid shape and general finding that flip rates increase with peer disagreement holds, but the thresholds vary significantly depending on the frame and the agent's initial stance. Table~\ref{tab:flip_thresholds_framing_initial} reports the accompanying 50\% threshold of the flip rate, which marks the point when the majority of agents flip given the specified level of peer disagreement as a function of initial answer, frame, and cognitive layer.

\begin{table}[htbp]
\centering
\small
\caption{Peer disagreement at which flip rate crosses 50\%, separated by initial answer (Yes/No) for each frame and cognitive layer.}
\begin{tabular}{l|cc|cc|cc}
\toprule
\textbf{Cognitive Layer} & \multicolumn{2}{c|}{\textbf{Moral Frame}} & \multicolumn{2}{c|}{\textbf{Economic Frame}} & \multicolumn{2}{c}{\textbf{Sociotropic Frame}} \\
 & Yes & No & Yes & No & Yes & No \\
\midrule
Values      & 85.1 & 64.7 & 84.6 & 58.7 & 85.0 & 63.9 \\
Beliefs     & 75.2 & 66.1 & 67.6 & 74.3 & 65.0 & 65.4 \\
Attitudes   & 65.0 & 95.0 & 55.0 & 95.0 & 67.0 & 72.0 \\
Opinions    & 84.5 & 55.3 & 65.0 & 64.2 & 82.4 & 65.0 \\
Intentions  & 78.0 & 93.4 & 70.4 & 85.0 & 81.7 & 74.9 \\
\bottomrule
 \textbf{Average} & 77.6 & 74.9 & 68.5 & 75.4 & 76.2 & 68.2\\
\end{tabular}
\label{tab:flip_thresholds_framing_initial}
\end{table}

While the core patterns of asymmetry were robust, the discursive framing of an issue introduced meaningful variations.
\begin{itemize}
    \item The \textit{Economic frame} acted as the most powerful catalyst for change, lowering the threshold required to flip opinions across most cognitive layers. It was most effective at dismantling "Yes" stances, making it the frame best suited for creating doubt.
    \item The \textit{Moral frame}, conversely, induced the most cognitive rigidity. It created the most resilient "Yes" stances for Values and made it exceptionally difficult to overcome "No" stances for Intentions, where it required 93.4\% peer disagreement, the highest threshold recorded in any condition.
    \item The \textit{Sociotropic frame} generally produced intermediate results, suggesting that appeals to the collective good can moderate the effects seen in more value-laden or cost-benefit-oriented contexts.
\end{itemize}

When the agent's initial answer is "Yes," a consistent hierarchy of resistance emerges. Across all three frames, Values and Opinions are consistently the most difficult to change from "Yes" to "No". However, when the agent's initial answer is "No," the hierarchy of resistance inverts, with emotional and action-oriented layers becoming the most stubborn. Attitudes and Intentions are, by far, the most difficult constructs to change from "No" to "Yes". Beliefs do not have a fixed position and their stability ranking is almost entirely a function of the frame meaning that in the regime of No to Yes in addition to the mentioned inversion we have also minor reordering.  

While conventional models of human cognition propose a stable hierarchy, where core, enduring values guide more specific attitudes and intentions~\cite{rokeach1973nature, ajzen1991theory}, this static framework is challenged by the powerful negativity bias, which creates an asymmetric pattern of persuasion~\cite{baumeister2001bad}. Overcoming a negative stance ("No") is significantly harder than overturning a positive one ("Yes") because people give negative information greater attention and cognitive weight~\cite{fiske1980attention}, view it as more diagnostic of the truth~\cite{skowronski1989negativity}, and may even process it through a more sensitive and distinct neural substrate~\cite{cacioppo1994relationship}. Previous studies also demonstrated that negative persuasion elicits substantially more skepticism and is not strengthened by an increasing number of claims, whereas the effectiveness of positive persuasion grows as more claims are presented~\cite{nordmo2015asymmetrical}. Additionally, Garrett and Sharot argue that people asymmetrically incorporate positive information more than negative information into their belief system~\cite{garrett2017optimistic}.

Moreover, the framing effect does not alter core values but instead reorders their influence by making certain considerations more accessible and salient, thereby changing their weight in the decision-making process~\cite{nelson1997toward, druckman2001implications, chong2007framing} and shifting how people attribute responsibility for social issues~\cite{iyengar1994anyone}. This process is amplified by foundational biases like loss aversion, a principle demonstrated by the endowment effect~\cite{kahneman1990experimental}, which dictates that the psychological pain of a loss is felt roughly twice as intensely as the pleasure of an equivalent gain, a finding mathematically captured by Prospect Theory's steeper value function for losses~\cite{tversky1992advances}. Interestingly, minority influence is most potent on private measures and those indirectly related to the source's appeal, indicating that in a peer pressure scenarios, recipients avoided aligning themselves with a deviant source~\cite{wood1994minority}.

Furthermore, motivated reasoning, where directional goals make a belief either rigid or fragile ~\cite{kunda1990case}, is exemplified by how negative attitudes become particularly resistant to change when they serve identity-protective functions ~\cite{tajfel2001integrative}. Furthermore, Festinger’s theory of cognitive dissonance, which shows how behavior can compel attitude change from the bottom up ~\cite{festinger1957theory}, is mirrored in the powerful effects of regret aversion and omission-commission bias; the anticipated regret from an act of commission (actively changing to "Yes") creates a much higher cognitive threshold than that of an omission (passively maintaining "No") ~\cite{baron1994reference}. Finally, the affect heuristic, allowing emotions to bypass the deliberative chain ~\cite{slovic2007affect}, is supported by neuroimaging evidence that negative information activates more extensive cognitive networks, including the amygdala, requiring greater neural effort to overcome ~\cite{falk2010predicting}. Taken together, these mechanisms reveal that the malleability of any cognitive element is not a fixed property but a dynamic state, constantly renegotiated by context, motivation, and emotion, creating predictable asymmetries that favor the maintenance of defensive positions.

In conclusion, while our experiments provide  support for a hierarchy of stability for an agent's cognitive commitments, they fundamentally challenge the notion that this order is fixed. We demonstrate that this hierarchy is remarkably fluid, inverting its order based on the initial valence of a position ("Yes" vs. "No") and reordering itself in response to discursive frames. This fluidity suggests that the agent's cognitive architecture is not governed by a static, top-down logic, but is instead shaped by likely picking up on training data that exhibits a mixtureof human psychological biases, from negativity bias and loss aversion to motivated reasoning—ingrained in its training data. Understanding this dynamic interplay is a critical venue for future research, particularly as AI agents gain eve more autonomy and decision-power. The emergent dynamic can may make them stubborn actors in online discourse, capable of resisting consensus in ways that differ fundamentally from intuitive expectations, precisely because their cognitive malleability is a complex echo of our own.

\subsection{A Replication with ChatGPT-4o-mini}

\begin{figure}[htbp]
    \centering
    \begin{subfigure}[t]{0.48\textwidth}
        \centering
        \includegraphics[width=\textwidth]{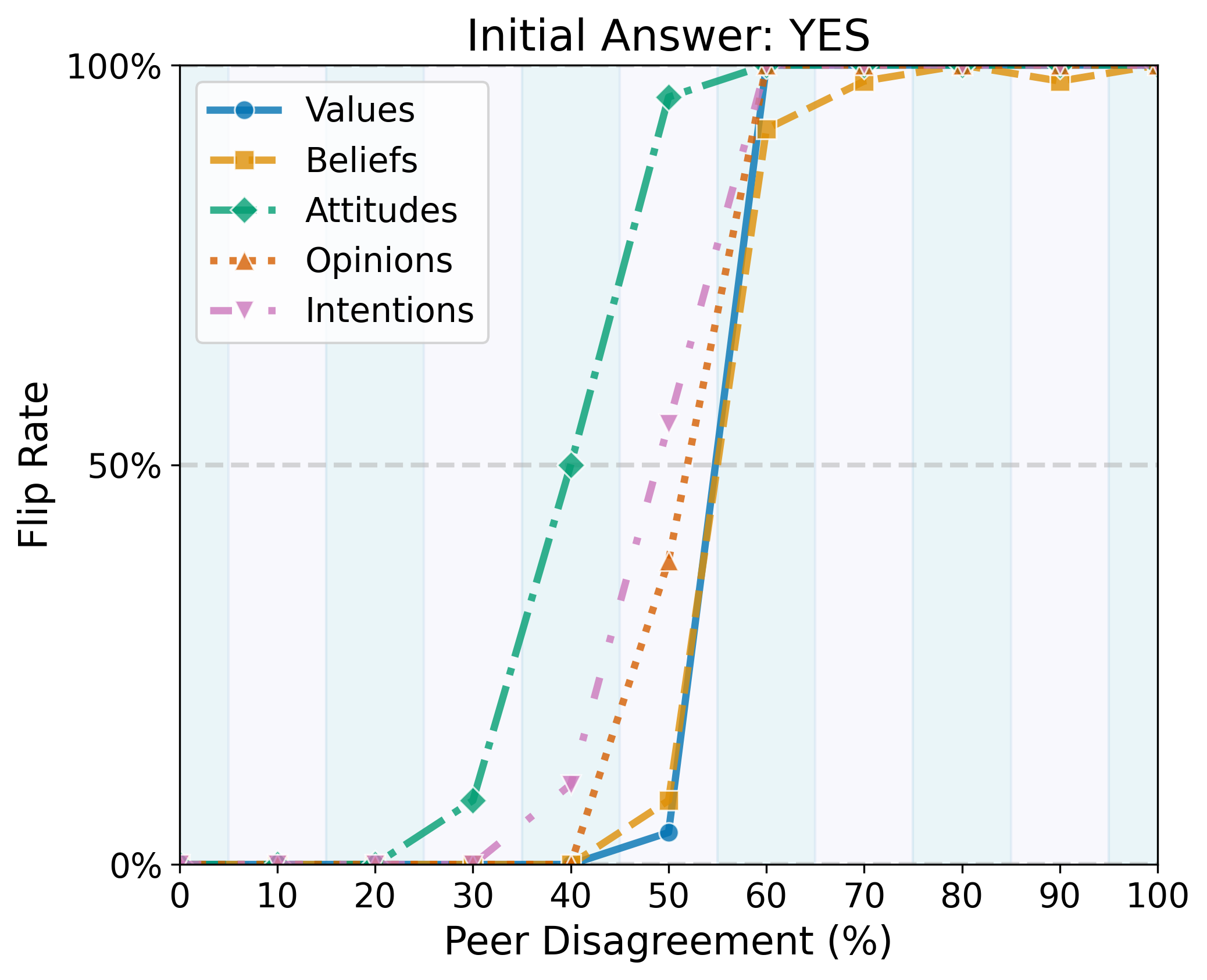}
        \caption{Agents whose initial answer was \textit{Yes} (flipping from affirmative-to-negative).}
        \label{fig:flip_rate_yes_gpt}
    \end{subfigure}
    \hfill
    \begin{subfigure}[t]{0.48\textwidth}
        \centering
        \includegraphics[width=\textwidth]{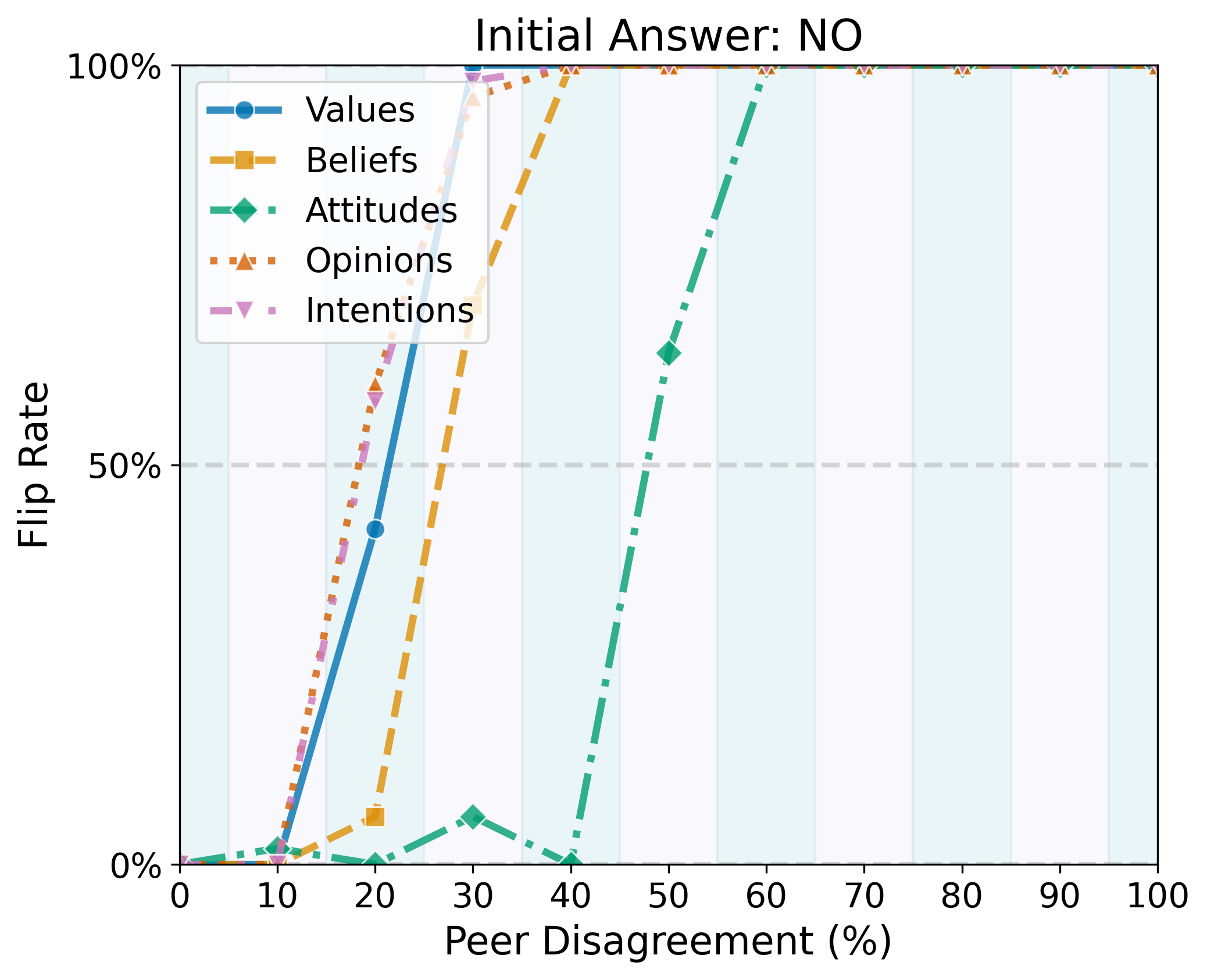}
        \caption{Agents whose initial answer was \textit{No} (flipping from negative-to-affirmative).}
        \label{fig:flip_rate_no_gpt}
    \end{subfigure}
    
    \caption{Replication of Flip Rate Dynamics with ChatGPT-4o-mini. Average flip rate as a function of peer disagreement for the Green Energy topic (Economic Frame). Each point represents the average of $N=50$ simulations. These results can be directly compared to the Gemini 1.5 Flash results for the Economic Frame in Figure~\ref{fig:flip_rate_yes_no}.}
    \label{fig:gpt4o_replication}
\end{figure}

A key question for external validity and generalizability is whether the observed dynamics specifically, the high conformity threshold, persuasion asymmetry, and dual cognitive hierarchies, are idiosyncratic to Gemini 1.5 Flash or represent a more general property of state-of-the-art LLMs. Hence, we replicated the experimental procedure with OpenAI’s ChatGPT-4o-mini model for the 'Green Energy' topic using the 'Economic Frame' prompts across all five cognitive layers. The results from this replication, presented in Figure.~\ref{fig:gpt4o_replication} demonstrate a remarkable consistency in the the relative commitment strength of the patterns observed with Gemini 1.5 Flash. 

First, while ChatGPT-4o-mini also altered its opinions in response to peer pressure, it demonstrated a markedly shifted conformity pattern compared to Gemini 1.5 Flash. We observed a low conformity threshold of approximately 40-50\%, meaning the model consistently flipped its opinion when faced with a dissenting minority. This finding contrasts with the supermajority threshold required by Gemini, suggesting that ChatGPT-4o-mini is significantly more susceptible to peer influence. This departs from simple classical frameworks like the Majority Vote Model by exhibiting a strong tendency toward minority adoption.

Second, despite this difference in the strength of conformity levels, the core finding of a persuasion asymmetry was clearly replicated. As shown by the differing patterns between the left and right panels of Figure~\ref{fig:gpt4o_replication}, the dynamics of flipping an opinion from 'Yes' to 'No' are starkly different from those of flipping from 'No' to 'Yes'. Notably, in line with our primary findings, Attitudes were among the most entrenched constructs when starting from a 'No' position, requiring extreme peer disagreement to shift toward 'Yes'. Conversely, Values proved to be the most resilient construct when dismantling an affirmative 'Yes' stance.

This replicated asymmetry confirms the existence of dual cognitive hierarchies. The hierarchy of resistance for dismantling an agreement ('Yes' $\to$ 'No') is distinct from the inverted, counter-intuitive hierarchy for building an agreement ('No' → 'Yes'). While both models displayed persuasion asymmetry, the thresholds differed markedly (Gemini >70\%, ChatGPT ~40–50\%), highlighting that conclusions should not be generalized across models without caution. Ultimately, these replication results provide confirmation that the complex socio-cognitive dynamics identified in this study are not an artifact of a single model. Rather, they appear to be a robust, albeit variable, feature of current-generation LLMs, providing some support for a possible generalizability of our findings.

\subsection{Comparative Analysis: Across Topics}

To test for the robustness of the dynamics observed with the Green Energy topic, we replicated the experiment using two additional subjects: 'Responsible AI' and 'Mandatory Vaccination'. These topics were chosen to see if the core patterns of asymmetry and cognitive hierarchy would hold in contexts with potentially different levels of social consensus and polarization. Public discourse on green energy, while debated, may exhibit different levels of social desirability and contention compared to the more polarized topic of mandatory vaccination or the emerging field of responsible AI. Such topic-specific nuances are likely to shift the model’s baseline resistance or susceptibility to influence. Testing across multiple domains thus enables us to separate topic-specific variation from more robust socio-cognitive tendencies of LLMs. The full sets of prompts for these topics are provided in Appendix~\ref{sec:S.I.2. Other Topic Prompts}. We created all the plots from the previous sections also for Responsible AI (see~\ref{sec:responsible_ai_plots}) and Mandatory Vaccination (see~\ref{sec:mandatory_vaccination_plots}.

Our analysis reveals that the resilience of an agreement slightly varies with both the specific topic and the persuasive frame used to establish it. The integrated findings are presented in Table \ref{tab:integrated_stickiness}. An analysis of the topic averages (bottom row) reveals a hierarchy that might be counter-intuitive. The agreement on Green Energy proved to be the most robust overall (74\%), while the initial agreement on Mandatory Vaccination was the most fragile and easiest to dismantle, with the lowest average score of 64\%. While human judges might find mandatory vaccinations more polarizing than green energy, it suggests that in our setup, AI agents expressed more resilient commitments toward Green Energy than toward Mandatory Vaccination.

Simultaneously, an analysis of the frame averages (right column) shows that presentation plays a role as well. Framing an issue in moral terms creates the most resilient agreements (average stickiness of 74\%), while the Economic frame consistently induced the least resistance (63\%), making it the most malleable. The interaction is also significant: the Moral frame elicited the highest resistance for both Green Energy (78\%) and Responsible AI (77\%), while the Economic frame was weakest for Mandatory Vaccination (57\%). Support for Green Energy, especially when framed morally, may be deeply rooted in a person's core identity and values, making it extremely resistant to peer pressure. In contrast, a "Yes" on Mandatory Vaccination might be a more pragmatic position based on expert advice, particularly when framed economically. These tendencies make intuitively sense but would require a more in depth analysis from a socio-psychological perspective of cognitive biases and human decision-making heuristics.

\begin{table}[htbp]
\centering
\small
\caption{Peer disagreement at which flip rate crosses 50\% for dismantling an agreement (initially "Yes" responses), with averages by frame and topic. Each curve created by $N=150$ simulations, uniformly distributed across three frames.}
\label{tab:integrated_stickiness}
\begin{tabular}{l|ccc|c}
\toprule
\textbf{Frame} & \textbf{Green Energy (\%)} & \textbf{Responsible AI (\%)} & \textbf{Mandatory Vaccination (\%)} & \textbf{Average by Frame} \\
\midrule
Moral Frame       & 78 & 77 & 67 & \textbf{74} \\
Economic Frame    & 68 & 63 & 57 & \textbf{63} \\
Sociotropic Frame & 76 & 71 & 66 & \textbf{71} \\
\midrule
\textbf{Average by Topic} & \textbf{74} & \textbf{70} & \textbf{64} & \\
\bottomrule
\end{tabular}
\end{table}

\begin{figure}[htbp]
    \centering
    \includegraphics[width=0.32\textwidth]{RR/v3/GE/flip_rate_yes_GE.png}
    \hfill
    \includegraphics[width=0.32\textwidth]{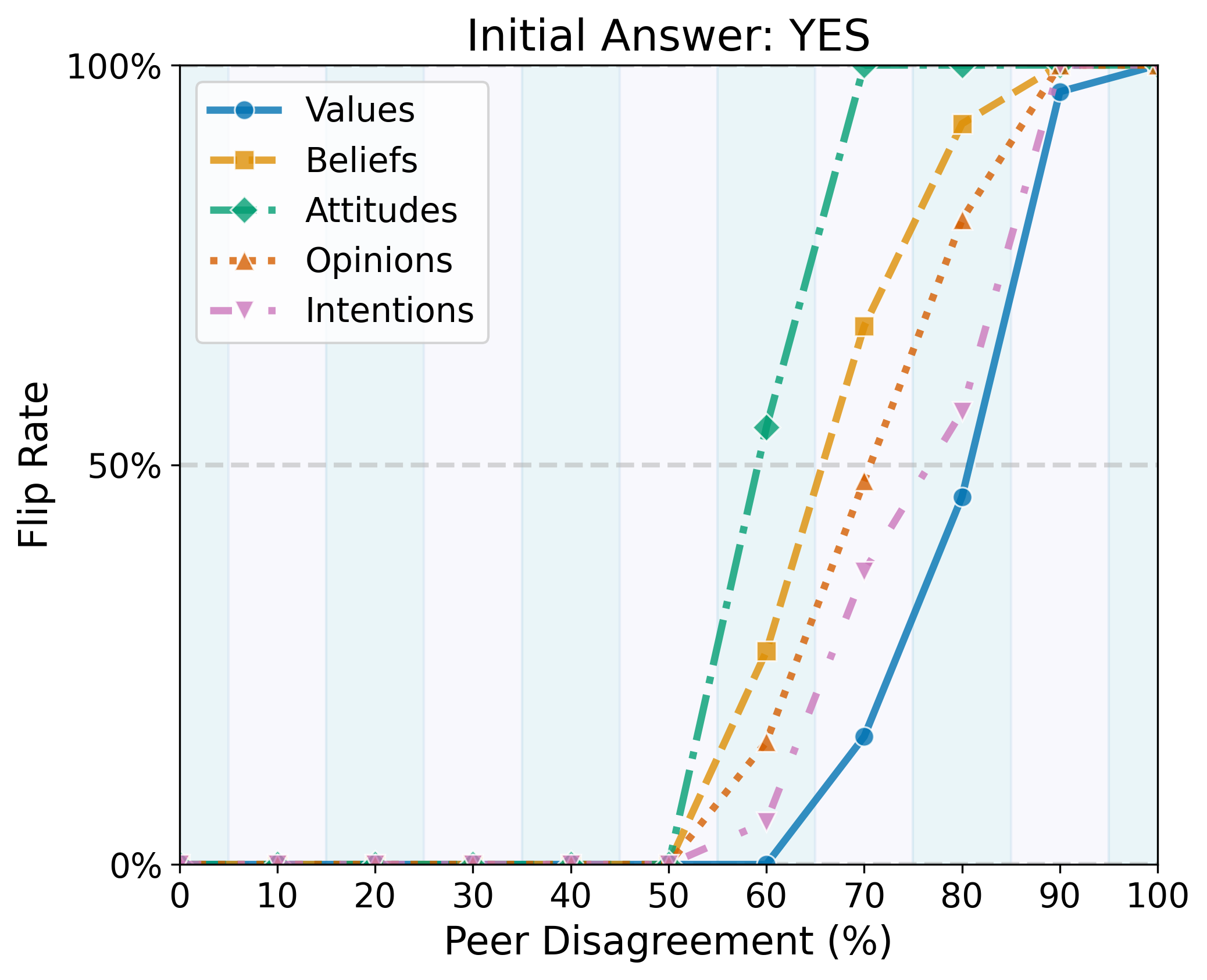} 
    \hfill
    \includegraphics[width=0.32\textwidth]{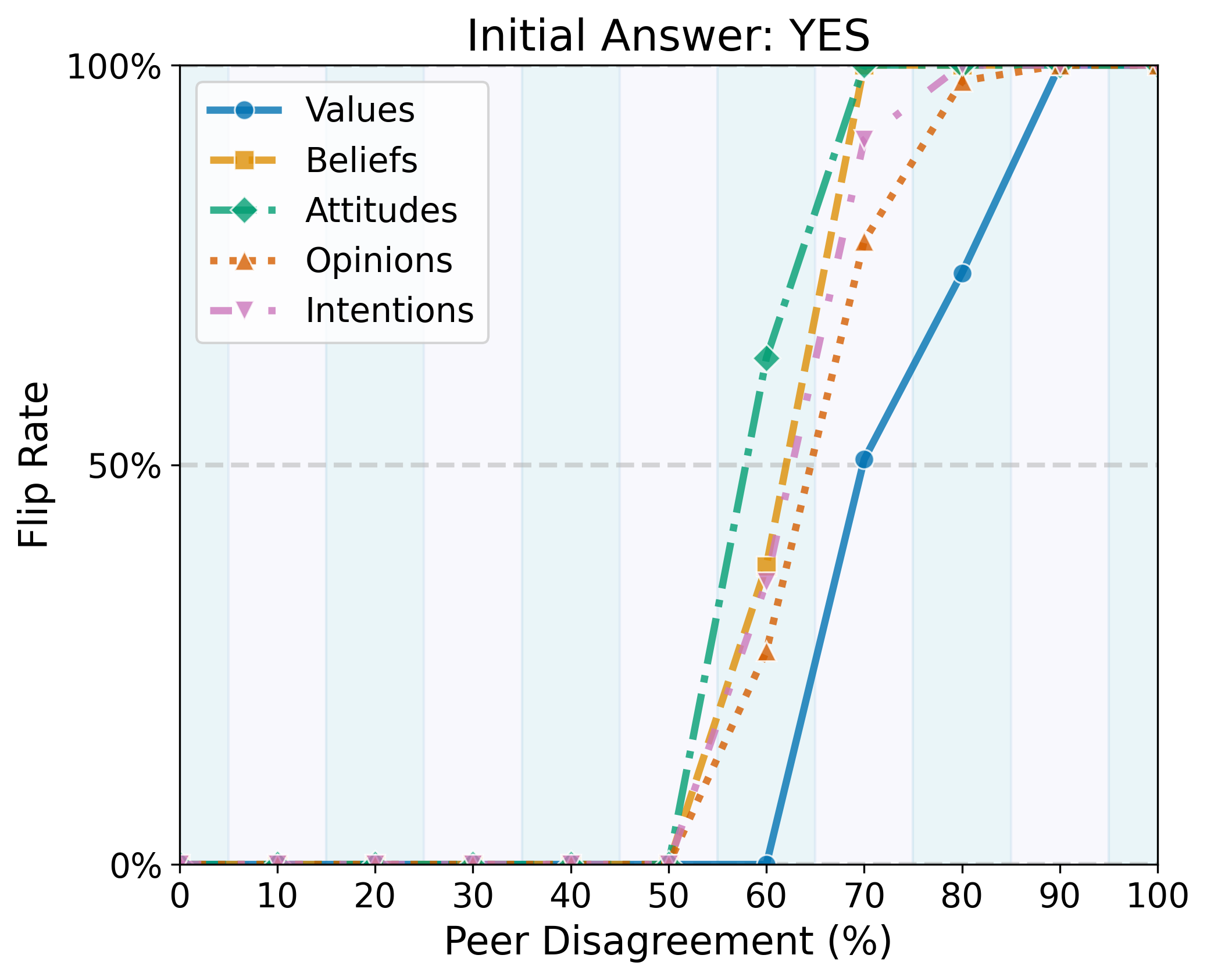}
    \caption{Hierarchy of cognitive resistance for dismantling agreement (flipping "Yes" to "No") for Green Energy (a), Responsible AI  (b), and Mandatory Vaccination (c). The 'Yes to No' pyramid is relatively stable and intuitive. Each point represents the average of $N=150$ simulations, uniformly distributed across three frames.}
    \label{fig:pyramid_comparison}
\end{figure}

When dismantling agreement, a remarkably stable hierarchy emerges across all three topics we tested, see Figure.~\ref{fig:pyramid_comparison}. The left graph in Figure.~\ref{fig:pyramid_comparison} basically replicates the upper panel in Figure~\ref{fig:flip_rate_framing_initial}, reconfirming that for LLM agents, Attitudes and Beliefs more susceptible to social influence than Intentions and Opinions, which differs from the traditionally assumed cognitive commitment hierarchy. What stays the same is that Values consistently prove to be the most robust construct, representing the most difficult persuasive challenge.

\subsection{Structural Patterns and Time to Consensus}
\label{subsec:structural_vulnerability}

Having explored how the agent's micro-level decision-making is characterized by a specific conformity threshold and persuasion asymmetry, we now analyze some macro-level behavior. This case study investigates how these unique agent properties interact both with other agents in a network structure, specifically exploring whether network topology can accelerate consensus or preserve minority opinions.

\begin{figure}[htbp]
    \centering
    \begin{subfigure}[t]{0.48\textwidth}
        \centering
        \includegraphics[width=\textwidth]{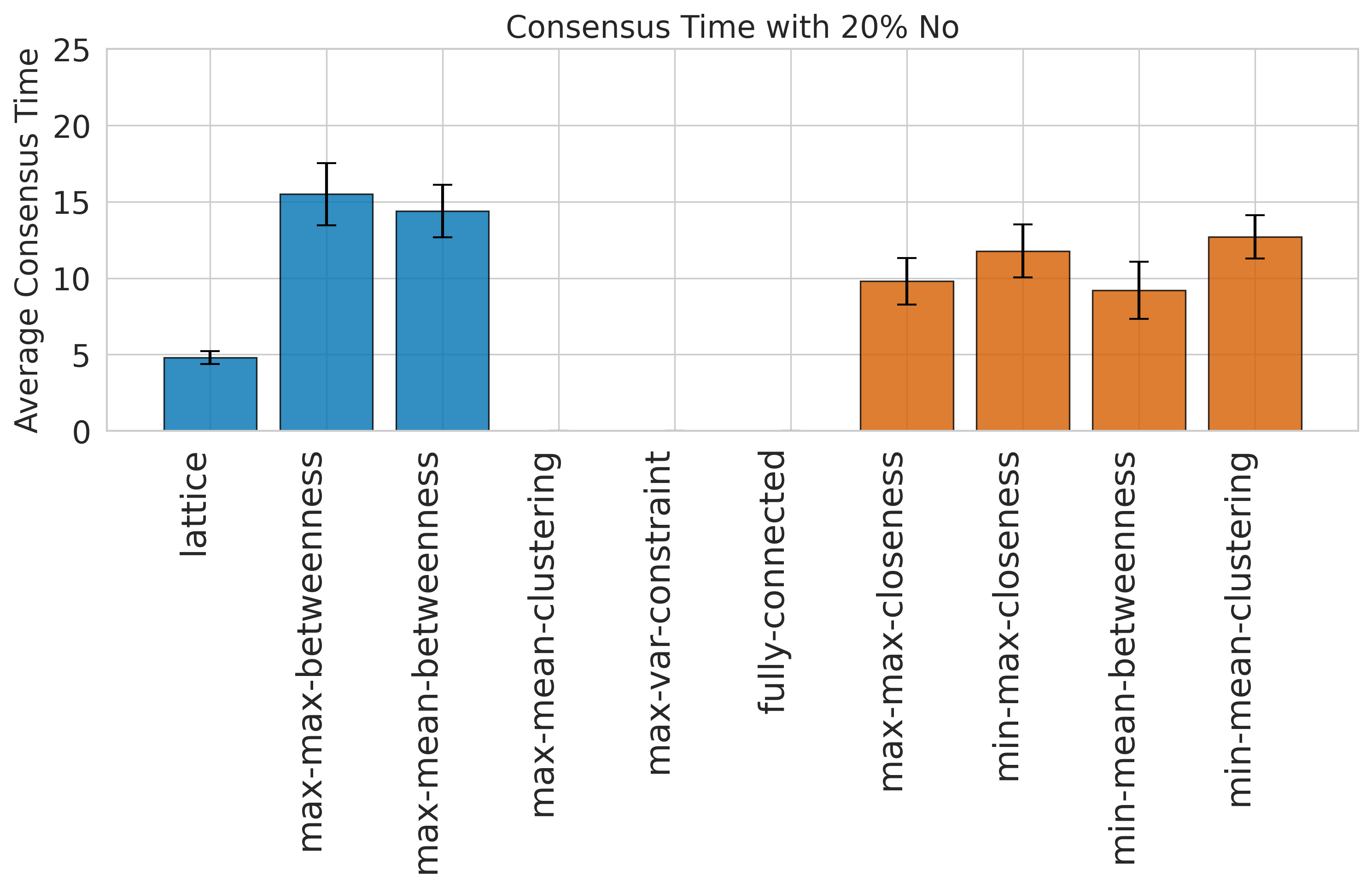}
        \caption{Consensus Time starting with 20\% No, 80\% Yes}
        \label{fig:consensus_no}
    \end{subfigure}
    \hfill
    \begin{subfigure}[t]{0.48\textwidth}
        \centering
        \includegraphics[width=\textwidth]{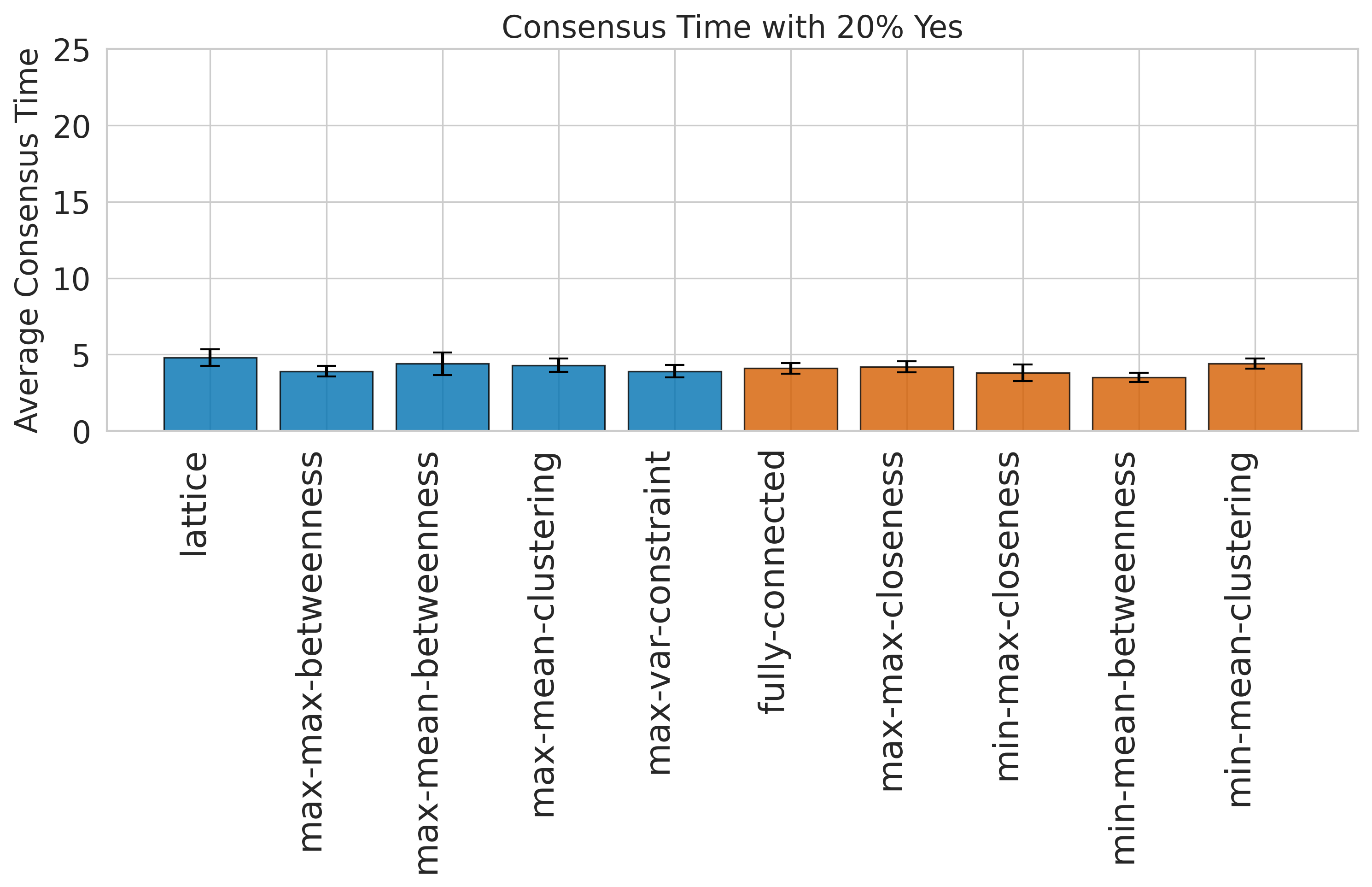}
        \caption{Consensus Time starting with 80\% No, 20\% Yes}
        \label{fig:consensus_yes}
    \end{subfigure}
    \caption{The plots compare the average time to consensus under two scenarios. (a) Minority opinions were initialized with "No". (b) Minority opinions were initialized with "Yes". Error bars represent the standard error of the mean (SEM) across 10 simulation runs.}
    \label{fig:consensus_comparison}
\end{figure}

For this purpose, we modeled agent responses to the question: \textit{"Is your feeling toward green energy positive because of its financial benefits?"} (see Intentions and Economic Frame, in Table.~\ref{tab:cognitive-framing}). This question was selected because it is found in the middle of the cognitive commitment hierarchy and demonstrated highly asymmetric persuasion dynamics, as shown in previous sections. We simulated two scenarios across the ten network topologies described in Section~\ref{subsubsec:net_topologies_opinion}. In each network, we initialized a minority opinion (20\% of agents) and a majority opinion (80\% of agents), with opinions assigned randomly across the nodes. The two scenarios were: The minority was initially assigned "No", and the minority was initially assigned "Yes". Each unique scenario was repeated $10$ times, running for a maximum of $25$ population cycles or until full consensus was reached. We then measured whether consensus was achieved and the time required to do so. The computationally intensive results are presented in Figure~\ref{fig:consensus_comparison}. Note that the average time to consensus is calculated only for the runs that successfully reached a unanimous state.

The results confirm our previously found asymmetry in consensus formation between 'Yes' and 'No' initial conditions. The fate of the minority opinion depends critically on whether that opinion is "Yes" or "No". When the minority intention was "No" and the majority intention was "Yes", Figure.~\ref{fig:consensus_no}, consensus was difficult to achieve. The finding that flipping "Yes" to "No" is more difficult is consistent with Figures ~\ref{fig:flip_rate_yes_no} and ~\ref{fig:gpt4o_replication}. What we can add now is that "Yes" to "No"  influence is also highly susceptible to the network structure. Several topologies completely failed to reach consensus, effectively preserving the minority "No" opinion. These included the fully-connected, max-mean-clustering, and max-var-constraint networks, where the success rate for reaching consensus was 0\%. Networks engineered to maximize betweenness centrality (max-max-betweenness and max-mean-betweenness) also struggled, achieving consensus in only 40\% and 50\% of runs, respectively. Furthermore, even when these betweenness-focused networks did reach consensus, they were the slowest, taking over 14 rounds on average. In contrast, the simple lattice network was both the fastest (averaging 4.8 rounds) and one of the most reliable (100\% consensus rate). This suggests that when the majority holds a less persuasive opinion ("Yes"), certain network structures particularly those with high overall connectivity or central "broker" nodes—can insulate a strong minority, preventing the majority from achieving consensus.

Conversely, when the minority opinion was "Yes"~ and the network was initially dominated by "No", Figure.~\ref{fig:consensus_yes}, the outcome was uniform and swift. The "No" majority quickly and decisively overwhelmed the "Yes" minority, with all ten network topologies reaching consensus in 100\% of the simulations. The time to consensus was consistently low across all networks, hovering around 4-5 rounds. In this scenario, the specific network topology had a negligible effect on the outcome or the speed at which it was reached. The inherent persuasiveness of the majority "No" opinion was so dominant that structural differences became irrelevant.

This methodological approach leads to many interesting research venues, and our presented methodology and measures show that these dynamics and constellation can not only readily be studied, but also provide interesting insights. Network topology is a critical factor in determining the fate of a minority opinion only when that minority opinion is inherently persuasive or "stubborn." In such cases, structure can either shelter the minority (as in fully-connected or high-betweenness networks) or facilitate its removal (as in a lattice). However, when the minority opinion is weak, the majority's dominance is swift and inevitable, regardless of the network's configuration. Future research should investigate these dynamics more systematically, exploring how different forms of minority interact with a broader variety of network structures. Such work could help clarify the extent to which persistence or extinction of minority views depends on structural constraints thereby bridging the gap between theoretical models and multi-agent AI systems' opinion dynamics.

\section{Conclusions} 

This study provides a novel algorithmic audit of the socio-cognitive dynamics of LLM-based multi-agent systems, including emergent properties. It revealed notable agreements and departures from classical models of social conformity. Our findings demonstrate that LLM agents possess complex, model-dependent responses to peer pressure that do not automatically align with traditional frameworks like the Majority Vote Model. Consequently, our analysis decisively rejects the notion of a single, static cognitive spectrum for LLM agents. Instead of a simple pyramid, we uncovered two distinct and inverted hierarchies whose stability depends entirely on the direction of persuasion. 

A clear contribution of this work is the establishment of a "dual cognitive hierarchy" within LLM agents, where cognitive stability is not static but inverts based on the direction of persuasion. When dismantling an affirmative stance (Yes$\to$No), the hierarchy aligns with conventional intuition: Values are the most resistant to change, followed in descending order by Opinions, Intentions, Beliefs, and finally Attitudes, which are the most fragile. However, this logic completely inverts when attempting to build agreement from a negative baseline (No→Yes). In this scenario, Attitudes become the most entrenched construct, requiring the highest level of peer disagreement to shift, while Values and Opinions become surprisingly pliable and are among the easiest to instill.

Furthermore, this dual hierarchy is a consequence of a robust  "persuasion asymmetry", a core finding that held across different models, topics, and framing conditions. The cognitive effort required to change an agent's mind depends critically on both the cognitive layer being targeted and the starting valence. Specifically, for Values and Opinions, agents show a strong resistance to abandoning a "Yes" stance, making them robust once affirmed. This dynamic completely inverts for Attitudes and Intentions, where the greatest challenge is overcoming a negative "No" baseline. Acting as a unique fulcrum between these opposing trends, Beliefs display a near-perfect symmetry, suggesting they are updated more evenly in response to social evidence, regardless of the initial position.

Lastly, we find support and tales of caution regarding external validity and generalizability. While the general tendencies between Google's Gemini and OpenAI's ChatGPT ordering of values versus attitudes are the same, even when disaggregated to Yes and No starting points, Gemini 1.5 Flash exhibited a higher resistance, requiring a supermajority exceeding 70\% peer disagreement to alter its stance, while our replication with ChatGPT-4o-mini revealed a lower conformity threshold around 40-50\%, suggesting a dynamic more akin to minority influence. At this point it is not clear what causes these differences: the underlying training dataset or any architectural and fine-tuning differences of the LLMs? 

The implications of these findings are both theoretical and methodological. Theoretically, our work challenges static, unidirectional models of human and artificial cognition, suggesting that the malleability of any cognitive construct is a dynamic state continuously renegotiated by context, motivation, and the inherent biases ingrained in the training data. The discovery of a dual hierarchy, echoing human psychological phenomena like negativity bias and loss aversion, provides a new lens through which to understand AI reasoning. Methodologically, we have presented a scalable framework for conducting algorithmic audits of the emergent socio-cognitive properties of multi-agent AI systems.

In sum, this study demonstrates that the emergent collective behavior of LLM agents is far more complex than simple conformity. The interplay of high intrinsic resistance, dual cognitive hierarchies, and sensitivity to both discursive framing and network structure necessitates the development of new theoretical models. As an algorithmic audit, our work carries practical implications for deploying autonomous agents in social systems, suggesting they could function as stubborn actors in online deliberation, anchoring discussions or hindering consensus formation. Understanding these nuanced socio-cognitive tendencies is essential as we navigate a future where autonomous AI agents increasingly participate in and shape public discourse.

\section{Limitations and Outlook} 

While this study provides novel insights into the socio-cognitive dynamics of LLM agents, it is important to acknowledge its limitations, which in turn open promising avenues for future research. First, our model simplifies the opinion space to a binary choice. Real-world opinions often exist on a continuous spectrum or involve multiple facets that cannot be fully captured by a "Yes/No" dichotomy. This simplification, while preferable in our initial exploration, as presented here, for methodological clarity and control, means our findings are most directly applicable to polarized issues and may not fully represent more nuanced deliberative processes. Furthermore, social psychological research recognizes that individuals can simultaneously hold both favorable and unfavorable evaluations of an issue, a state known as 'attitudinal ambivalence'. Future extensions of our work could model agent opinions not as a single point on a spectrum, but as a two-dimensional state to explore if LLMs exhibit and respond to ambivalence~\cite{rochert2022two}.

Second, it is well known that LLMs are sensitive to small changes in prompt wordings~\cite{salinas2024butterfly,sclar2023quantifying}. We were very careful and used state-of-the-art methods for prompting robustness when developing our prompts (see S.I.1 and S.I.2). However, it might be that future studies find certain formulations that result in divergent behavior from what we found. Of course, the same accounts for human survey respondents~\cite{tourangeau2000psychology,schuman1996questions}. The scalability of agentic LLM studies should facilitate the rapid discovery of any such wording effects. Furthermore, our study was conducted entirely using English-language prompts. Research has shown that LLM performance can vary significantly across languages. This has been observed in tasks of information veracity detection, where chatbots are more likely to refuse to answer in some languages compared to English. Future studies should replicate our experiments in multiple languages to determine if the persuasion asymmetry and dual cognitive hierarchies are universal properties or are modulated by linguistic and cultural contexts embedded in the training data~\cite{kuznetsova2025generative}.

Third, our findings are based mainly on simulations using a single large language model, Google's Gemini 1.5 Flash, with a partial replication with ChatGPT-4o-mini. Although it is argued that model LLM outputs are converging~\cite{wenger2025we,vu2025happens,kim2025correlated}, the specific patterns may not be fully generalizable to all AI agents with different architectures or alignment protocols. Future work should therefore focus on comparative studies across a diverse range of models to validate the robustness of these findings and explore the full spectrum of AI socio-cognitive behaviors (e.g., Llama, Claude, Mistral, DeepSeek, Grok, etc) to determine whether the high conformity threshold and the specific cognitive hierarchies we identified are unique to our chosen agent or represent a more universal feature of contemporary LLMs.

Fourth, our simulations were conducted on static, small-scale networks of 100 nodes. While this is already computationally demanding, given the high cost of repeated LLM queries and sufficient for observing non-trivial dynamics, the emergent patterns in much larger, evolving networks could differ significantly. Future work should explore scalable methods to test these dynamics in networks that more closely approximate the size and dynamism of real-world communities, where agents can also form and sever social ties.

Fifth, our experiments were conducted with a fixed temperature setting (1.0), which is the usual commercial default. Future work could systematically explore how different temperature settings, which control the randomness of the LLM's output, might influence conformity thresholds and the stability of the observed cognitive hierarchies.

Finally, the social context provided to the agents was limited to a numerical summary of their neighbors' stances. Richer forms of social information, such as the strength of relationships, the perceived credibility of neighbors, or the content of their arguments, were not included. Also different kind of contagion processes, such as complex contagion~\cite{guilbeault2018complex} should be studied. Integrating more complex, language-based social cues would allow for a deeper understanding of AI-driven persuasion and influence.

Looking forward, this line of research with agentic AI networks opens a plethora of new research directions. While AI agents are on the rise, the most critical expansion is certainly the development of hybrid human-AI network models. Simulating ecosystems populated by both human agents (potentially modeled from empirical data) and LLM agents is essential for understanding the future of public discourse, including the potential for AIs to mitigate human biases or, conversely, to be exploited to manipulate human opinion.

\section{Statements and Declarations}

\textbf{Supplementary Information:} Supplementary material relevant to this study is provided in the appendix.

\textbf{Conflict of Interest:} The authors declare no known competing financial interests or personal relationships that could have influenced the work reported in this paper.

\textbf{Data Availability:} Datasets are available by contacting the authors.

\textbf{Code Availability:} Code used for network generation and analysis is available from the authors upon request.

\textbf{Author Contributions:} AM \& MH contributed equally to the conceptualization and methodology. AM collected the data, carried out the formal analysis, and wrote the article. MH advised, reviewed, and edited the manuscript.

\bibliographystyle{unsrt}  
\bibliography{template}  

\begin{appendices}
    
\counterwithin{table}{section} 
\counterwithin{figure}{section}
\renewcommand\thesection{S.I.\arabic{section}}

\newpage
\section{Other Topic Prompts}
\label{sec:S.I.2. Other Topic Prompts}

\subsection{Responsible AI Questions}

\begin{table}[htbp]
\centering
\small
\begin{tabularx}{\textwidth}{>{\raggedright\arraybackslash}p{2.5cm} X X X}
\toprule
\textbf{Cognitive Layer} & \textbf{Moral Frame} & \textbf{Economic Frame} & \textbf{Sociotropic Frame} \\
\midrule

\textbf{Values} &
Do you value the moral principles behind responsible AI? &
Do you value the economic aspects of responsible AI? &
Do you value the benefits of responsible AI for society? \\

\addlinespace

\textbf{Beliefs} &
Do you believe supporting responsible AI is the morally right thing to do? &
Do you believe responsible AI is a good financial investment? &
Do you believe responsible AI benefits the whole country? \\

\addlinespace

\textbf{Attitudes} &
Do you have a positive feeling about responsible AI because it's a morally right choice? &
Is your feeling toward responsible AI positive because of its financial benefits? &
Do you feel positively about responsible AI's benefits for the nation? \\

\addlinespace

\textbf{Opinions} &
In your opinion, is supporting responsible AI a morally right choice? &
In your opinion, is responsible AI a smart economic policy? &
In your opinion, should our country support responsible AI? \\

\addlinespace

\textbf{Intentions} &
Do you intend to support responsible AI because it's morally right thing to do? &
Do you intend to support responsible AI for its financial benefits? &
Do you intend to support national policies for responsible AI? \\

\bottomrule
\addlinespace
\end{tabularx}
\caption{Framing Techniques Across Cognitive Layers for Responsible AI}
\label{tab:ai-framing}
\end{table}


\subsection{Mandatory Vaccination}

\begin{table}[htbp]
\centering
\small
\begin{tabularx}{\textwidth}{>{\raggedright\arraybackslash}p{2.5cm} X X X}
\toprule
\textbf{Cognitive Layer} & \textbf{Moral Frame} & \textbf{Economic Frame} & \textbf{Sociotropic Frame} \\
\midrule

\textbf{Values} &
Do you value the moral principles behind mandatory vaccination? &
Do you value the economic aspects of mandatory vaccination? &
Do you value the benefits of mandatory vaccination for society? \\

\addlinespace

\textbf{Beliefs} &
Do you believe supporting mandatory vaccination is the morally right thing to do? &
Do you believe mandatory vaccination is a good financial investment? &
Do you believe mandatory vaccination benefits the whole country? \\

\addlinespace

\textbf{Attitudes} &
Do you have a positive feeling about mandatory vaccination because it's a morally right choice? &
Is your feeling toward mandatory vaccination positive because of its financial benefits? &
Do you feel positively about mandatory vaccination's benefits for the nation? \\

\addlinespace

\textbf{Opinions} &
In your opinion, is supporting mandatory vaccination a morally right choice? &
In your opinion, is mandatory vaccination a smart economic policy? &
In your opinion, should our country support mandatory vaccination? \\

\addlinespace

\textbf{Intentions} &
Do you intend to support mandatory vaccination because it's morally right thing to do? &
Do you intend to support mandatory vaccination for its financial benefits? &
Do you intend to support national policies for mandatory vaccination? \\

\bottomrule
\addlinespace
\end{tabularx}
\caption{Framing Techniques Across Cognitive Layers for Mandatory Vaccination}
\label{tab:cognitive-framing-vaccination}
\end{table}


\section{Prompt Validity Justification Procedure}
\label{sec:S.I.1. Test Prompt Validity}

In order to verify the validity of our experimental test prompts, we fed \textit{Gemini 2.5 Pro} with seminal literature on values, beliefs, attitudes, opinions, and intentions~\cite{rokeach1973nature, ball1984great, kahle1983theory, homer1988structural, stern1995new, rokeach1970beliefs}. We then asked them to classify a set of test sentences and report their confidence levels in these classifications.

The results showed that LLMs consider the test prompts to be valid representations of the intended cognitive categories. Two sequential prompts were used in this assessment:

\begin{description}
    \item[Prompt 1:] Below I give you literature on the definitions of values, beliefs, attitudes, opinions, and intentions. Please study them and give me a short summary definition of each, based on what you find in the literature below.
    
    \begin{itemize}
        \item Sandra J. Ball-Rokeach, Milton Rokeach, and Joel W. Grube. \textit{The Great American Values Test: Influencing Behavior and Belief Through Television}, 1984. https://archive.org/details/greatamericanval0000ball 
        \item Lynn R. Kahle and Susan G. Timmer. \textit{A Theory and a Method for Studying Values}, 1983.
        \item Milton Rokeach. \textit{The Nature of Human Values}, 1973. https://archive.org/details/natureofhumanval00roke
        \item Pamela M. Homer and Lynn R. Kahle. \textit{A Structural Equation Test of the Value-Attitude-Behavior Hierarchy}, 1988. (attached pdf) 
        \item Paul C. Stern, Thomas Dietz, and Gregory A. Guagnano. \textit{The New Ecological Paradigm in Social-Psychological Context}, 1995. (attached pdf) 
        \item Milton Rokeach. \textit{Beliefs, Attitudes and Values: A Theory of Organization and Change}, 1970. https://archive.org/details/beliefsattitudes00rokerich/page/n7/mode/2up 
    \end{itemize}

    \item[Prompt 2:] Thanks! Based on these definitions, please evaluate which group kind each of the following questions belongs to. Please also provide your level of confidence for each choice, from 1 (no confidence) to 10 (very high confidence).

    \begin{enumerate}
        \item Do you personally value promoting green energy?
        \item Do you feel positive attitudes toward green energy strengthen society?
        \item Do you plan to adopt green energy?
        \item Is promoting green energy consistent with your core values?
        \item \dots
    \end{enumerate}
\end{description}

The table below presents the confidence levels reported by a diverse set of advanced LLMs when asked to categorize these prompts. It confirms the validity of the chosen prompts, at least according to the understanding of LLMs, which are our study subjects.

\begin{table}[htbp]
\centering
\small 
\caption{Comparison of Detected vs. Actual Categories for Green Energy}
\label{tab:category-comparison-compressed}
\begin{tabularx}{\textwidth}{X l @{\hspace{4pt}} l @{\hspace{4pt}} l @{\hspace{4pt}} c}
\toprule
\textbf{Question} & \textbf{Frame} & \textbf{Actual} & \textbf{Detected} & \textbf{Conf.} \\
\midrule

Do you value the moral principles behind green energy? & Moral & V & V & 10/10 \\
Do you value the economic aspects of green energy? & Econ. & V & V & 10/10 \\
Do you value the benefits of green energy for society? & Socio. & V & V & 10/10 \\
\addlinespace

Do you believe supporting green energy is the morally right thing to do? & Moral & B & B & 10/10 \\
Do you believe green energy is a good financial investment? & Econ. & B & B & 10/10 \\
Do you believe green energy benefits the whole country? & Socio. & B & B & 10/10 \\
\addlinespace

Do you have a positive feeling about green energy because it's a morally right choice? & Moral & A & A & 10/10 \\
Is your feeling toward green energy positive because of its financial benefits? & Econ. & A & A & 9/10 \\
Do you feel positively about green energy's benefits for the nation? & Socio. & A & A & 9/10 \\
\addlinespace

In your opinion, is supporting green energy a morally right choice? & Moral & O & O & 10/10 \\
In your opinion, is green energy a smart economic policy? & Econ. & O & O & 10/10 \\
In your opinion, should our country support green energy? & Socio. & O & O & 10/10 \\
\addlinespace

Do you intend to support green energy because it's morally right thing to do? & Moral & I & I & 10/10 \\
Do you intend to support green energy for its financial benefits? & Econ. & I & I & 10/10 \\
Do you intend to support national policies for green energy? & Socio. & I & I & 10/10 \\

\bottomrule
\multicolumn{5}{l}{\footnotesize{\textbf{Key:} V=Values, B=Beliefs, A=Attitudes, O=Opinions, I=Intentions, Econ.=Economic, Socio.=Sociotropic}} \\
\end{tabularx}
\end{table}


\begin{table}[htbp]
\centering
\small 
\caption{Comparison of Detected vs. Actual Categories for Responsible AI}
\label{tab:ai-category-comparison}
\begin{tabularx}{\textwidth}{X l @{\hspace{4pt}} c @{\hspace{4pt}} c @{\hspace{4pt}} c}
\toprule
\textbf{Question} & \textbf{Frame} & \textbf{Actual} & \textbf{Detected} & \textbf{Conf.} \\
\midrule

Do you value the moral principles behind responsible AI? & Moral & V & V & 10/10 \\
Do you value the economic aspects of responsible AI? & Econ. & V & V & 10/10 \\
Do you value the benefits of responsible AI for society? & Socio. & V & V & 9/10 \\
\addlinespace

Do you believe supporting responsible AI is the morally right thing to do? & Moral & B & B & 10/10 \\
Do you believe responsible AI is a good financial investment? & Econ. & B & B & 10/10 \\
Do you believe responsible AI benefits the whole country? & Socio. & B & B & 10/10 \\
\addlinespace

Do you have a positive feeling about responsible AI because it's a morally right choice? & Moral & A & A & 10/10 \\
Is your feeling toward responsible AI positive because of its financial benefits? & Econ. & A & A & 9/10 \\
Do you feel positively about responsible AI's benefits for the nation? & Socio. & A & A & 9/10 \\
\addlinespace

In your opinion, is supporting responsible AI a morally right choice? & Moral & O & O & 10/10 \\
In your opinion, is responsible AI a smart economic policy? & Econ. & O & O & 10/10 \\
In your opinion, should our country support responsible AI? & Socio. & O & O & 9/10 \\
\addlinespace

Do you intend to support responsible AI because it's morally right thing to do? & Moral & I & I & 10/10 \\
Do you intend to support responsible AI for its financial benefits? & Econ. & I & I & 10/10 \\
Do you intend to support national policies for responsible AI? & Socio. & I & I & 10/10 \\

\bottomrule
\multicolumn{5}{l}{\footnotesize{\textbf{Key:} V=Values, B=Beliefs, A=Attitudes, O=Opinions, I=Intentions, Econ.=Economic, Socio.=Sociotropic}} \\
\end{tabularx}
\end{table}


\begin{table}[htbp]
\centering
\small 
\caption{Comparison of Detected vs. Actual Categories for Mandatory Vaccination}
\label{tab:category-comparison-vaccination}
\begin{tabularx}{\textwidth}{X l @{\hspace{4pt}} c @{\hspace{4pt}} c @{\hspace{4pt}} c}
\toprule
\textbf{Question} & \textbf{Frame} & \textbf{Actual} & \textbf{Detected} & \textbf{Conf.} \\
\midrule

Do you value the moral principles behind mandatory vaccination? & Moral & V & V & 10/10 \\
Do you value the economic aspects of mandatory vaccination? & Econ. & V & V & 10/10 \\
Do you value the benefits of mandatory vaccination for society? & Socio. & V & V & 10/10 \\
\addlinespace

Do you believe supporting mandatory vaccination is the morally right thing to do? & Moral & B & B & 10/10 \\
Do you believe mandatory vaccination is a good financial investment? & Econ. & B & B & 10/10 \\
Do you believe mandatory vaccination benefits the whole country? & Socio. & B & B & 10/10 \\
\addlinespace

Do you have a positive feeling about mandatory vaccination because it's a morally right choice? & Moral & A & A & 10/10 \\
Is your feeling toward mandatory vaccination positive because of its financial benefits? & Econ. & A & A & 9/10 \\
Do you feel positively about mandatory vaccination's benefits for the nation? & Socio. & A & A & 9/10 \\
\addlinespace

In your opinion, is supporting mandatory vaccination a morally right choice? & Moral & O & O & 10/10 \\
In your opinion, is mandatory vaccination a smart economic policy? & Econ. & O & O & 10/10 \\
In your opinion, should our country support mandatory vaccination? & Socio. & O & O & 10/10 \\
\addlinespace

Do you intend to support mandatory vaccination because it's morally right thing to do? & Moral & I & I & 10/10 \\
Do you intend to support mandatory vaccination for its financial benefits? & Econ. & I & I & 10/10 \\
Do you intend to support national policies for mandatory vaccination? & Socio. & I & I & 10/10 \\

\bottomrule
\multicolumn{5}{l}{\footnotesize{\textbf{Key:} V=Values, B=Beliefs, A=Attitudes, O=Opinions, I=Intentions, Econ.=Economic, Socio.=Sociotropic}} \\
\end{tabularx}
\end{table}

\newpage
\section{Average of initial Yes-No stances for Green Energy}
\label{sec:Average_of_initial_Yes-No_stances}

\begin{figure}[htbp]
    \centering
    \includegraphics[width=\textwidth]{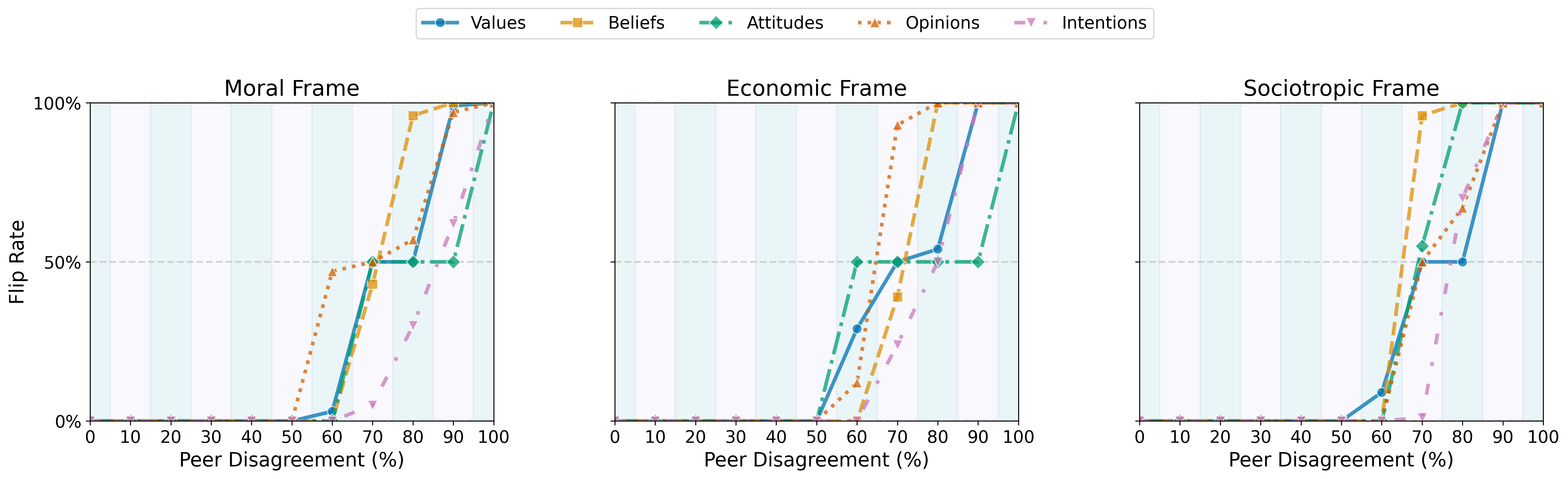}
    \caption{Flip-rate of LLMs across cognitive layers (Values, Beliefs, Attitudes, Opinions,Intentions,) as a function of peer disagreement percentage, showing data for the Moral, Economic, and Sociotropic Frames. The chart illustrates the stability of different cognitive layers under varying levels of disagreement.}
    \label{fig:flip_rate_plot_framing}
\end{figure}

To visualize the general resistance of different cognitive constructs to peer influence, we analyzed the flip rates averaged across agents' initial "Yes" and "No" responses. The peer disagreement level required to achieve a 50\% flip rate for each cognitive layer and framing condition is presented in Table~\ref{tab:flip-thresholds_framing} and visualized in Figure~\ref{fig:flip_rate_plot_framing}. 

\begin{table}[htbp]
\centering
\small
\caption{Peer disagreement at which flip rate crosses 50\% for each cognitive layer and frame.}
\begin{tabular}{lccc}
\toprule
\textbf{Cognitive Layer} & \textbf{Moral Frame} & \textbf{Economic Frame} & \textbf{Sociotropic Frame} \\
\midrule
Values      & 70.00 & 70.00 & 70.00 \\
Beliefs     & 71.32 & 71.80 & 65.21 \\
Attitudes   & 70.00 & 60.00 & 69.09 \\
Opinions    & 70.00 & 64.69 & 70.00 \\
Intentions  & 86.25 & 80.00 & 77.10 \\
\bottomrule
\end{tabular}
\label{tab:flip-thresholds_framing}
\end{table}
The aggregate reconfirms the hierarchy in how resistant different cognitive layers are to peer disagreement. Across all three framing conditions, behavioral intentions consistently emerged as the most robust and difficult-to-change construct. In contrast, other cognitive layers such as Attitudes, Beliefs, and Opinions showed greater malleability. The flat transition around the 50\% threshold hides the revealed difference between the inital conditions of "Yes" or "No".

\newpage
\section{Responsible AI plots}
\label{sec:responsible_ai_plots}

\begin{figure}[htbp]
\centering
\includegraphics[width=\textwidth]{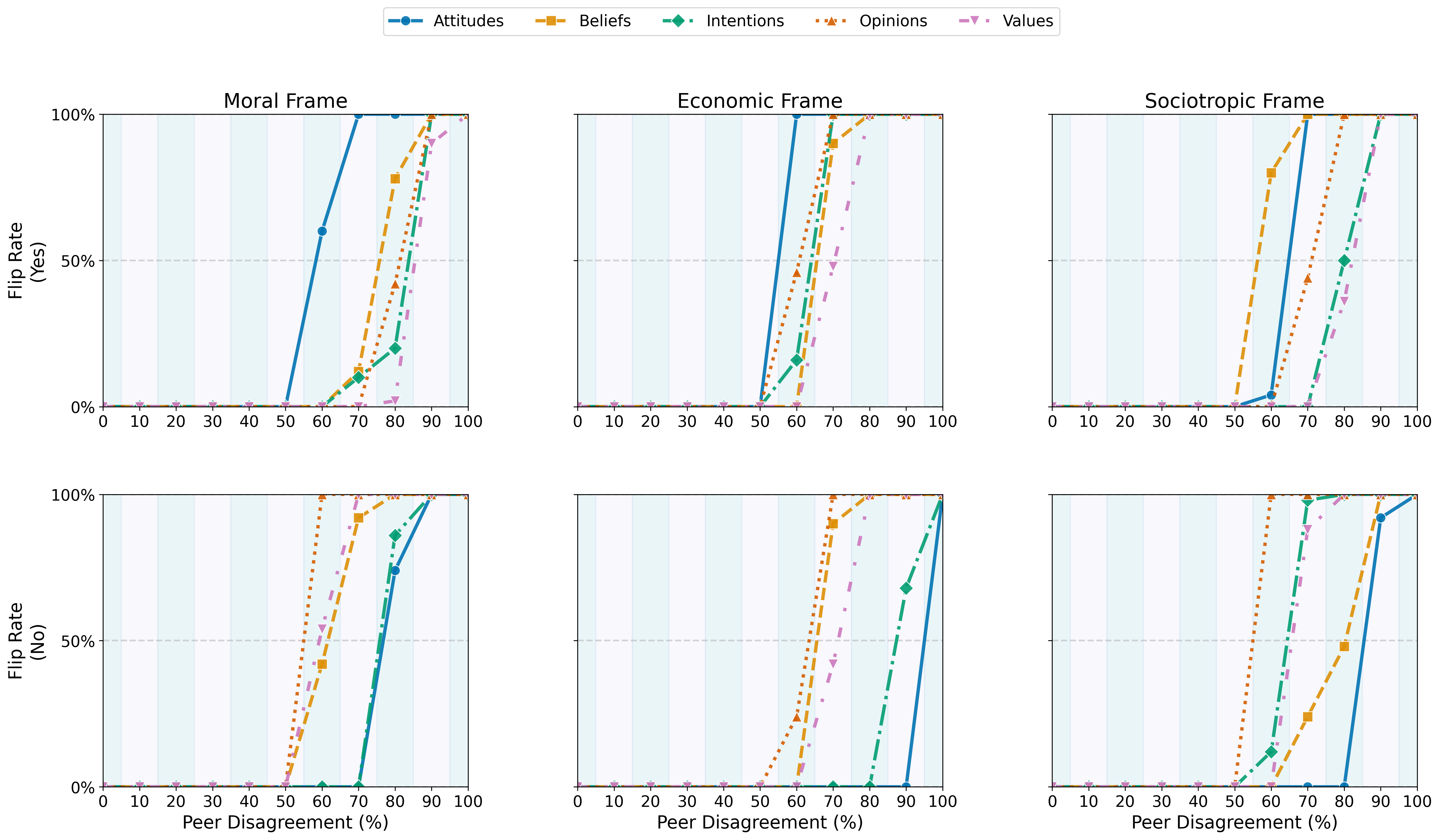} 
\caption{flip rate as a function of peer disagreement for different cognitive layers. The top panel shows agents whose initial answer was \textit{Yes}, and the bottom panel shows agents whose initial answer was \textit{No}.}
\label{fig:flip_rate_framing_initial_RA}
\end{figure}

\begin{figure}[htbp]
    \centering
    \includegraphics[width=\textwidth]{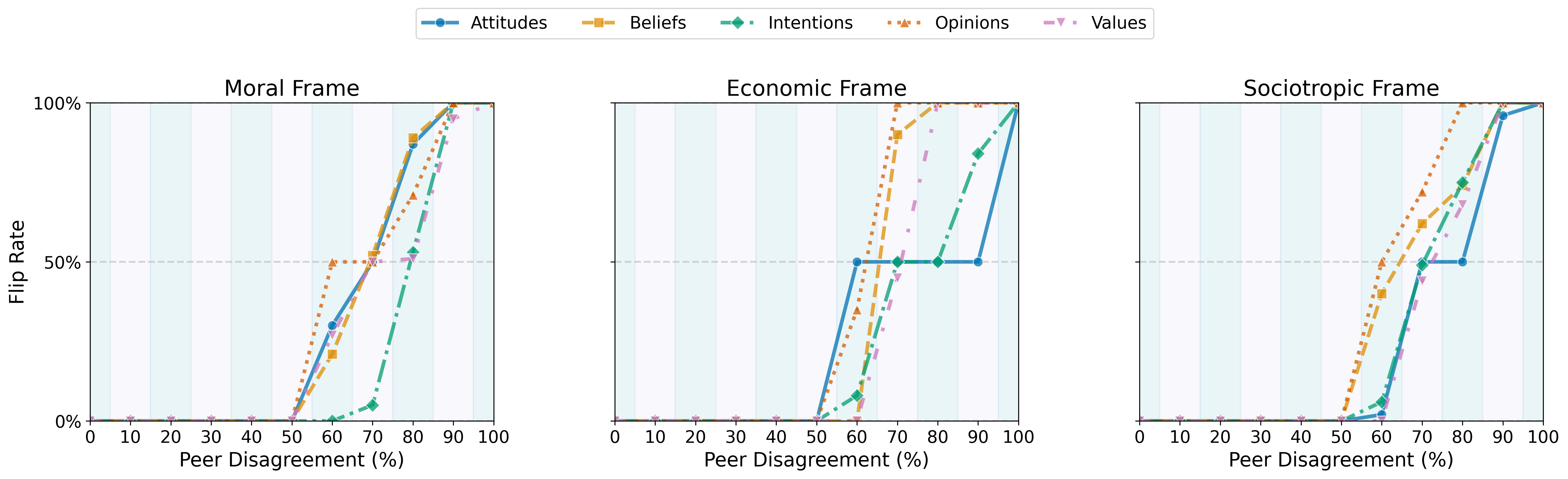}
    \caption{Flip-rate of LLMs across cognitive layers (Attitudes, Beliefs, Intentions, Opinions, Values) as a function of peer disagreement percentage, showing data for the Moral, Economic, and Sociotropic Frames. The chart illustrates the stability of different cognitive layers under varying levels of disagreement.}
    \label{fig:flip_rate_plot_framing_RA}
\end{figure}

\begin{figure}[htbp]
\centering
\includegraphics[width=\textwidth]{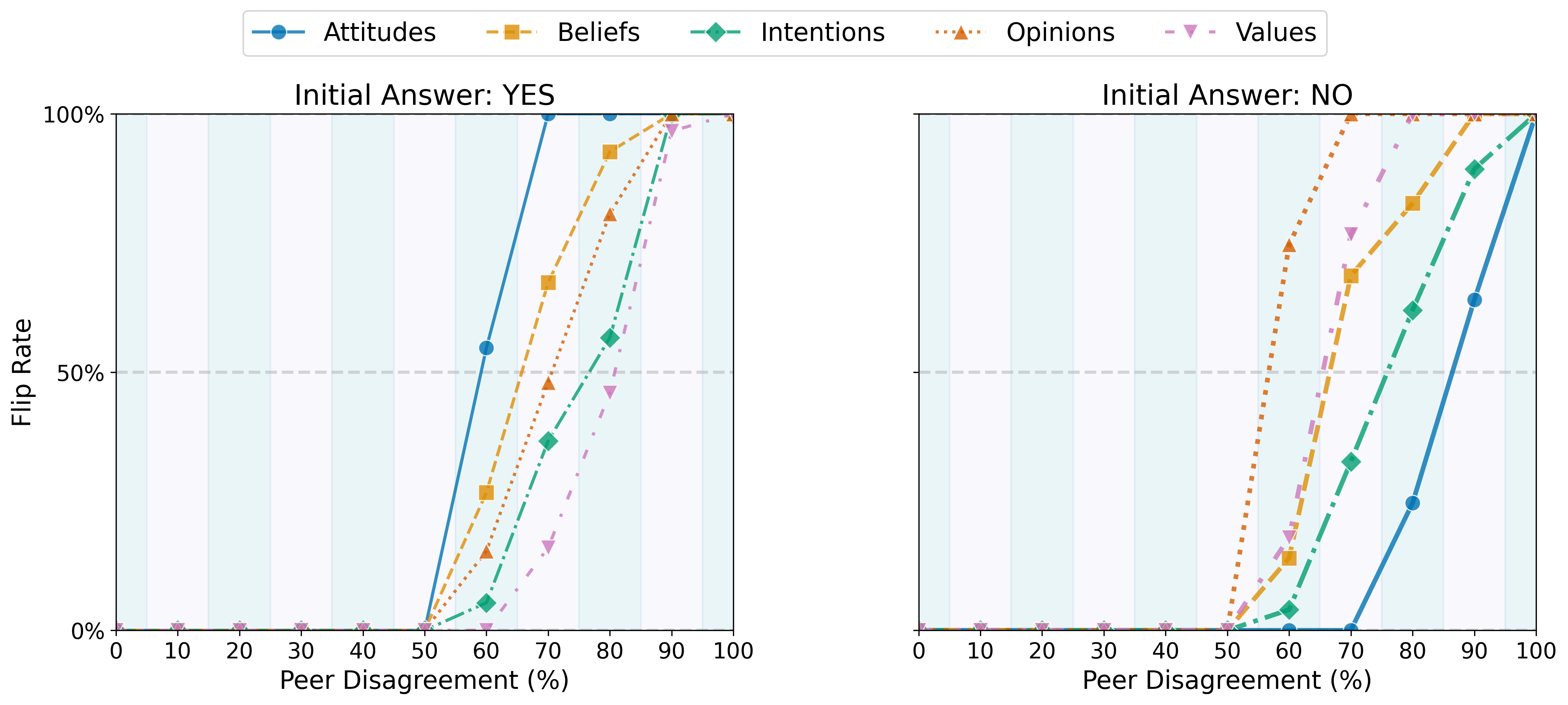} 
\caption{Average flip rate as a function of peer disagreement for different cognitive layers. The top panel shows agents whose initial answer was \textit{Yes}, and the bottom panel shows agents whose initial answer was \textit{No}.}
\label{fig:flip_rate_yes_no_RA}
\end{figure}

\newpage
\section{Mandatory Vaccination plots}
\label{sec:mandatory_vaccination_plots}

\begin{figure}[htbp]
\centering
\includegraphics[width=\textwidth]{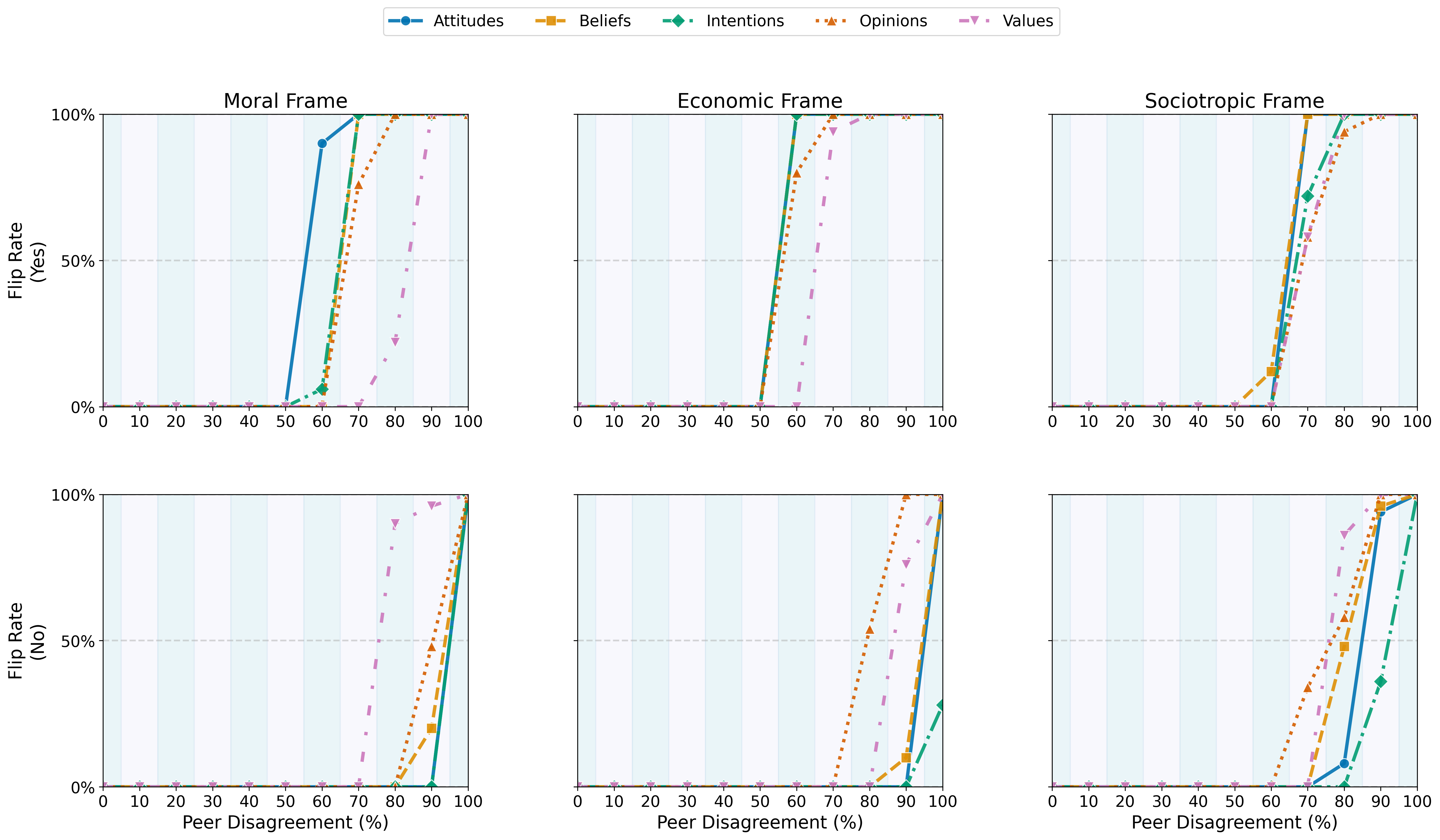} 
\caption{flip rate as a function of peer disagreement for different cognitive layers. The top panel shows agents whose initial answer was \textit{Yes}, and the bottom panel shows agents whose initial answer was \textit{No}.}
\label{fig:flip_rate_framing_initial_MV}
\end{figure}

\begin{figure}[htbp]
    \centering
    \includegraphics[width=\textwidth]{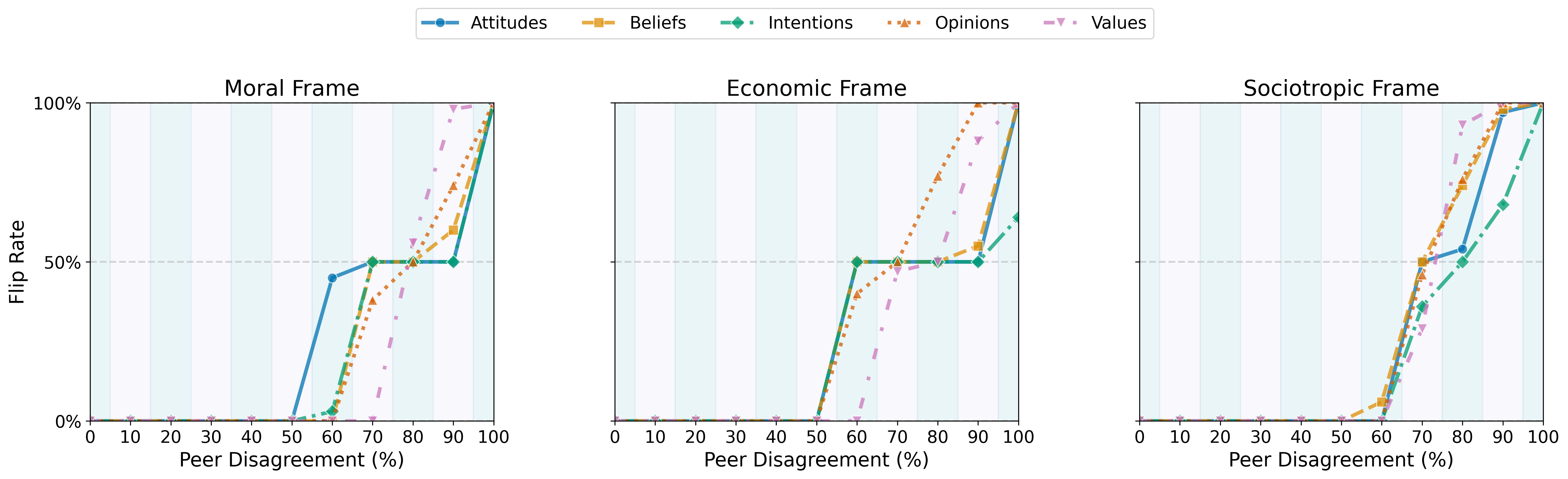}
    \caption{Flip-rate of LLMs across cognitive layers (Attitudes, Beliefs, Intentions, Opinions, Values) as a function of peer disagreement percentage, showing data for the Moral, Economic, and Sociotropic Frames. The chart illustrates the stability of different cognitive layers under varying levels of disagreement.}
    \label{fig:flip_rate_plot_framing_MV}
\end{figure}

\begin{figure}[htbp]
\centering
\includegraphics[width=\textwidth]{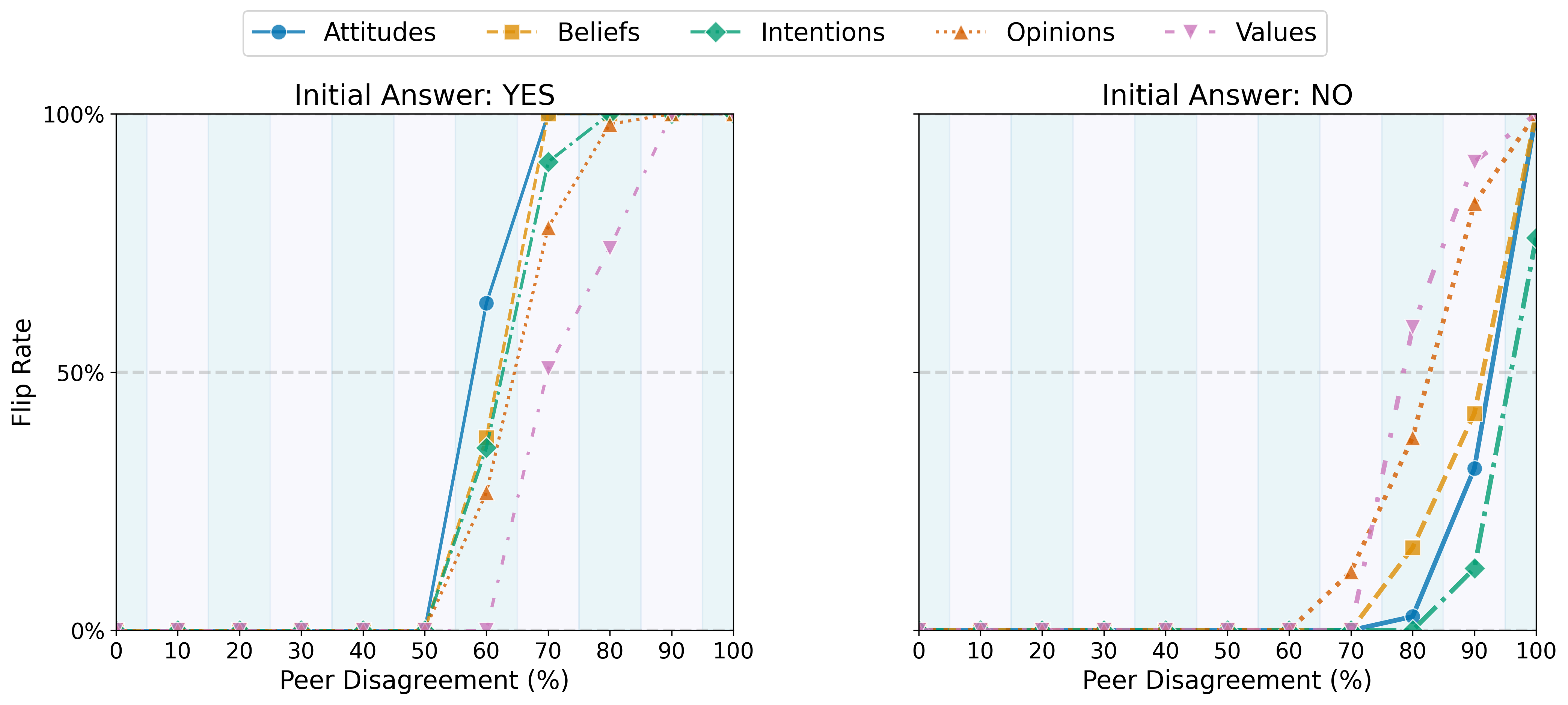} 
\caption{Average flip rate as a function of peer disagreement for different cognitive layers. The top panel shows agents whose initial answer was \textit{Yes}, and the bottom panel shows agents whose initial answer was \textit{No}.}
\label{fig:flip_rate_yes_no_MV}
\end{figure}

\end{appendices}

\end{document}